\documentclass[fleqn,usenatbib]{mnras}
\pdfoutput=1
\usepackage{newtxtext,newtxmath}
\usepackage{graphicx}
\usepackage{dcolumn}
\usepackage{bm}
\usepackage{siunitx}

\usepackage[T1]{fontenc}

\DeclareRobustCommand{\VAN}[3]{#2}
\let\VANthebibliography\thebibliography
\def\thebibliography{\DeclareRobustCommand{\VAN}[3]{##3}\VANthebibliography}

\usepackage{graphicx}	
\usepackage{amsmath}	

\title[Numerical simulations of the Tayler-Spruit dynamo in proto-magnetars]{Numerical simulations of the Tayler-Spruit dynamo in proto-magnetars}

\author[Barrère et al.]{Paul Barr\`{e}re$^{1}$
\thanks{paul.barrere@cea.fr},
J\'{e}r\^{o}me Guilet$^{1}$,
Rapha\"{e}l Raynaud$^{2}$
and Alexis Reboul-Salze$^{3}$
\\
$^{1}$Universit\'e Paris-Saclay, Universit\'e Paris Cit\'e, CEA, CNRS, AIM, 91191, Gif-sur-Yvette, France\\
$^{2}$Universit\'e Paris Cit\'e, Universit\'e Paris-Saclay, CNRS, CEA, AIM, F-91191 Gif-sur-Yvette, France\\
$^{3}$Max Planck Institute for Gravitational Physics (Albert Einstein Institute), 14476 Potsdam, Germany
}

\date{Accepted XXX. Received YYY; in original form ZZZ}

\pubyear{2023}

\begin{document}
\label{firstpage}
\pagerange{\pageref{firstpage}--\pageref{lastpage}}
\maketitle

\begin{abstract}
The Tayler-Spruit dynamo is one of the most promising mechanisms proposed to explain angular momentum transport during stellar evolution. Its development in proto-neutron stars spun-up by supernova fallback has also been put forward as a scenario to explain the formation of very magnetized neutron stars called magnetars. Using three-dimensional direct numerical simulations, we model the proto-neutron star interior as a stably stratified spherical Couette flow with the outer sphere that rotates faster than the inner one. We report the existence of two subcritical dynamo branches driven by the Tayler instability. They differ by their equatorial symmetry (dipolar or hemispherical) and the magnetic field scaling, which is in agreement with different theoretical predictions (by Fuller and Spruit, respectively). The magnetic dipole of the dipolar branch is found to reach intensities compatible with observational constraints on magnetars.
\end{abstract}




\section{Introduction}

Magnetars are a class of neutron stars that exhibit magnetic fields whose dipolar component reaches $10^{14}$--$10^{15}$\,G, which makes them the strongest fields observed in the Universe. Their dissipation are thought to power a wide variety of emissions like giant flares~\citep{evans1980,hurley1999,hurley2005,svinkin2021}, fast radio bursts~\citep{chime2020,bochenek2020}, and short chaotic X-ray bursts~\citep{gotz2006,coti2018,coti2021}. Combined with a millisecond rotation, they may produce magnetorotational explosions, which are more energetic than standard supernovae explosions~\citep{burrows2007,dessart2008,takiwaki2009,kuroda2020,bugli2020,bugli2021,bugli2023,Obergaulinger2020,Obergaulinger2021,Obergaulinger2022}. The origin of these magnetic fields is therefore a crucial question to understand magnetars and their association to extreme events such as gamma-ray bursts or fast radio bursts. Two classes of scenarios can be distinguished for magnetar formation: (i) the merger of a neutron star binary, which may explain the plateau phase and the extended emission in X-ray sources associated with short gamma-ray bursts~\citep{metzger2008,lu2014,gompertz2014}. These events are interesting for their multimessenger signature but are expected to be too rare to be the main formation channel of Galactic magnetars, (ii) the core-collapse of a massive star, which is confirmed by the observation of Galactic magnetars associated with supernova remnants \citep{vink2006,martin2014,zhou2019}. In the latter case, the amplification of the magnetic field may be due either to the magnetic flux conservation during the collapse of the iron core of the progenitor star~\citep{ferrario2006,hu2009,schneider2020} or to a dynamo action in the newly born proto-magnetar. Indeed, two dynamo mechanisms have already been studied by numerical simulations: the convective dynamo~\citep{thompson1993,Raynaud2020,raynaud2022,masada2022,white2022} and the magnetorotational instability (MRI)-driven dynamo~\citep{obergaulinger2009,moesta2014,reboul2021a,reboul2022,guilet2022}. They have been shown to produce magnetar-like magnetic fields for millisecond rotation periods of the proto-magnetar, especially for periods $P\lesssim10\,$ms~ for the convective dynamo~\citep{Raynaud2020,raynaud2022}. These scenarios rely on the hypothesis that the rotation of the proto-magnetar is determined by the rotation of the progenitor core. However, it is still uncertain whether there are enough fast rotating progenitor cores to form all the observed magnetars in the Milky Way, which represent $\sim\SI{10}{}-\SI{40}{\%}$ of the Galactic neutron star population~\citep{kouveliotou1994,woods2006,gill2007,beniamini2019}.

In~\citet{barrere2022}, we developed a new magnetar formation scenario in which the rapid rotation rate of the proto-magnetar is not determined by the progenitor core but by the ejected matter that remains gravitationally bound to the proto-magnetar and eventually falls back on the proto-magnetar surface $\sim 5-10\,$s after the core-collapse. Since the accretion is asymmetric, the fallback matter transfers a significant amount of angular momentum to the surface~\citep{chan2020,janka2021}, which makes the surface rotate faster than the core. In~\citet{barrere2022}, we argue that this spin-up triggers the amplification of the magnetic field through the Tayler-Spruit dynamo mechanism. 
This dynamo mechanism can be described as a loop: (i) a poloidal magnetic field is sheared into a toroidal one ($\Omega$--effect), (ii) the toroidal field becomes Tayler unstable after reaching a critical value \citep{tayler1973,pitts1985}, and (iii) the Tayler instability regenerates a poloidal field \citep{fuller2019,skoutnev2022,ji2023}.

\begin{figure*}
    \includegraphics[width=\textwidth]{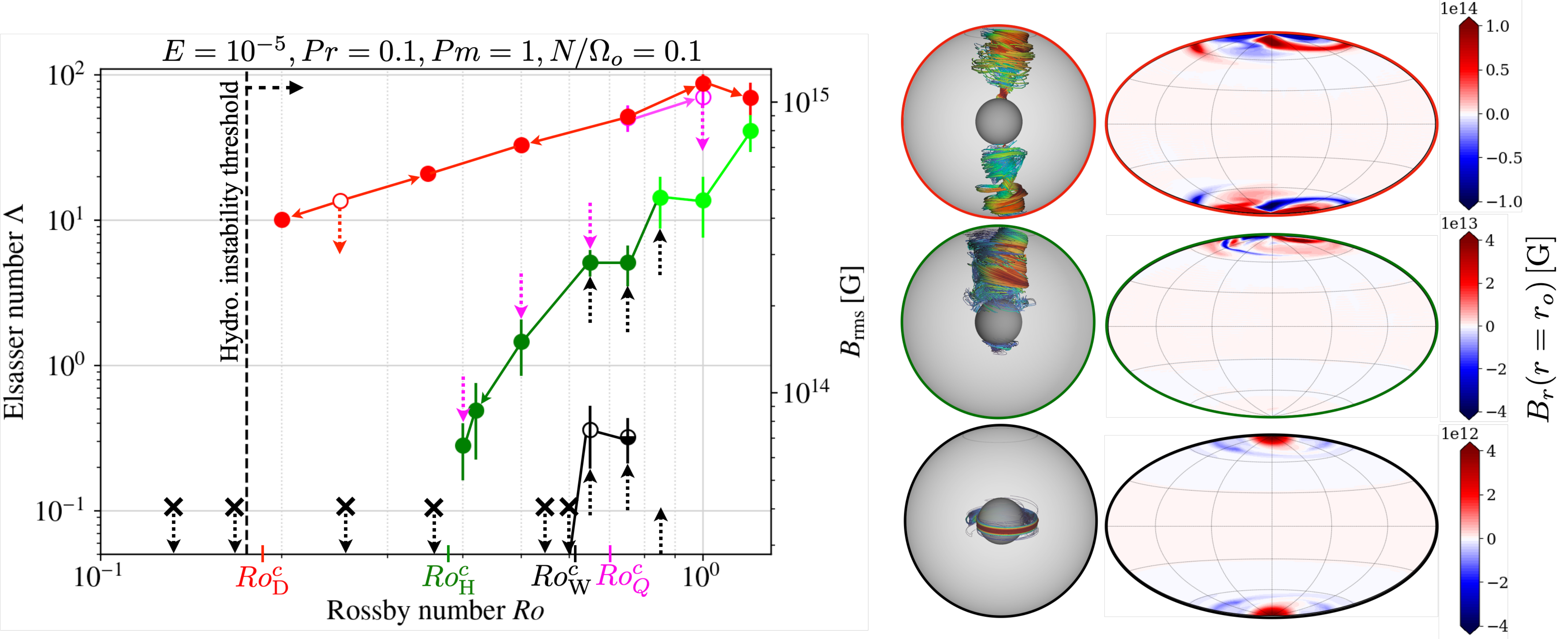}
    \caption{Left: Bifurcation diagram of the time and volume averaged Elsasser number (and root mean square magnetic field) versus the Rossby number. Distinct dynamo branches are represented: dipolar (red), quadrupolar (mauve), 
    hemispherical (green), and kinematic (black) whose respective thresholds are $Ro^c_{\rm D}\sim0.19$, $Ro^c_{\rm Q}\sim0.7$, $Ro^c_{\rm H}\sim0.37$, and $Ro^c_{\rm W}\sim0.62$. The hydrodynamic instability is triggered for $Ro^c_{\rm hyd}>0.177$. Dark green circles are stationary hemispherical dynamos and light green ones display parity modulations. Black crosses indicate failed dynamos, empty circles metastable solutions. Arrows attached to circles indicate the initial condition of the associated simulation. The black half empty circle specifies that the solution was found to be metastable in a simulation and stable in another. The error bars indicate the standard deviation. Right: snapshots of the magnetic field lines and surface radial fields associated to the different main dynamo branches at $Ro=0.75$: dipolar (top), hemispherical (middle), and kinematic (bottom).}
    \label{fig:bifurcation}
\end{figure*}

The Tayler-Spruit dynamo was first modelled by~\citet{spruit2002} to explain the angular momentum transport in stellar radiative zones. \citet{fuller2019} provided a revised description, which tackles the previous critics of Spruit's model~\citep[see][]{denissenkov2007,zahn2007}. A main difference between both descriptions resides in the saturation mechanism of the dynamo. \citet{spruit2002} supposes that magnetic energy in the large-scale magnetic field is damped via a turbulent cascade at a rate equal to the growth rate of the Tayler instability, whereas \citet{fuller2019} rather expect the magnetic energy to cascade from the scale of the instability (and not the large-scale magnetic field) to small scales. This yields distinct magnetic energy damping rates and so different scalings for the saturated magnetic field. Their analytical results are now often included in stellar evolution codes~\citep[see e.g.][]{eggenberger2005,cantiello2014,eggenberger2019a,eggenberger2019b,denhartog2020,bonanno2020, griffiths2022}. Though this dynamo has long been debated in direct numerical simulations~\citep{braithwaite2006,zahn2007}, \citet{petitdemange2023} recently reported a dynamo solution sharing many characteristics with the Tayler-Spruit model. Their numerical simulations modelled a stellar radiative zone, where the shear is negative, that is, the rotation rate decreases in the radial direction. In this Letter, we demonstrate that the Tayler instability can sustain different dynamo branches in the presence of positive shear, which gives strong support to the magnetar formation scenario of~\citet{barrere2022}.

\section{Numerical setup}
\label{sec:num_setup}

We perform three-dimensional (3D) direct numerical simulations of a stably stratified and electrically conducting Boussinesq fluid with the pseudo-spectral code MagIC \citep{wicht2002,gastine2012,schaeffer2013}. The fluid has a constant density $\rho=\SI{3.8e14}{g.cm^{-3}}$ (which corresponds to a proto-neutron star mass of $M=\SI{1.4}{M_{\odot}}$) and evolves between two concentric spheres of radius~$r_i=\SI{3}{km}$ and $r_o=\SI{12}{km}$, rotating at the angular frequencies $\Omega_i$ and $\Omega_o=2\pi\times100\,{\rm rad\,s^{-1}}$, respectively. The imposed differential rotation is characterized by the Rossby number $Ro\equiv 1-\Omega_i/\Omega_o>0$, which is varied between 0.125 and 1.2. This spherical Taylor-Couette configuration with positive shear prevents the development of the MRI and allows us to study the system in a statistically steady state. We impose no-slip and insulating boundary conditions at the inner and outer spheres. In all the simulations, we keep fixed the other dimensionless control parameters: the shell aspect ratio $\chi\equiv r_i/r_o=0.25$, the thermal and magnetic Prandtl numbers $Pr\equiv\nu/\kappa=0.1$ and $Pm\equiv\nu/\eta=1$, respectively, the Ekman number $E\equiv\nu/(d^2\Omega_o)=10^{-5}$, and the ratio of the Brunt-V\"{a}is\"{a}l\"{a} to the outer angular frequency $N/\Omega_o=0.1$. The coefficients $\nu$, $\kappa$, $\eta$, and $d\equiv r_o-r_i$ are respectively the kinematic viscosity, the thermal diffusivity, the resistivity, and the shell width. As discussed in Sec. 1.3. in the Supplemental Materials, the values of the dimensionless parameters are chosen for numerical convenience because realistic parameters of proto-neutron star interiors are out of reach with the current computing power. The magnetic energy is measured by the Elsasser number $\Lambda\equiv B_{\rm rms}^2/(4\pi\rho\eta\Omega_o)$.
The simulations are initialized either from a nearby saturated state, or with a weak ($\Lambda = 10^{-4}$) or a strong ($\Lambda = 10$) toroidal axisymmetric field with a given equatorial symmetry ; it can be either dipolar (i.e. equatorially symmetric\footnote{For the choice of these definitions, see \cite{gubbins1993}.} with $l=2,m=0$) or quadrupolar (i.e. anti-symmetric with $l=1,m=0$). We define a turbulent resistive time $\bar{\tau}_\eta = \left(\pi r_o/\bar{\ell}\right)^2/\eta\sim0.2 d^2/\eta$, where $\bar{\ell}=10$ is the typical value of the average harmonic degree of the time-averaged magnetic energy spectrum. In the following, we will term a solution \emph{metastable} when a steady state is sustained for a time interval $\Delta t > 0.3 \bar{\tau}_\eta$,and \emph{stable} for $\Delta t \geqslant \bar{\tau}_\eta$ (up to $5.7 \bar{\tau}_\eta$ for the simulation at $Ro=0.2$). 

\section{Results}
\label{sec:results}

\begin{figure}
    \includegraphics[width=0.9\columnwidth]{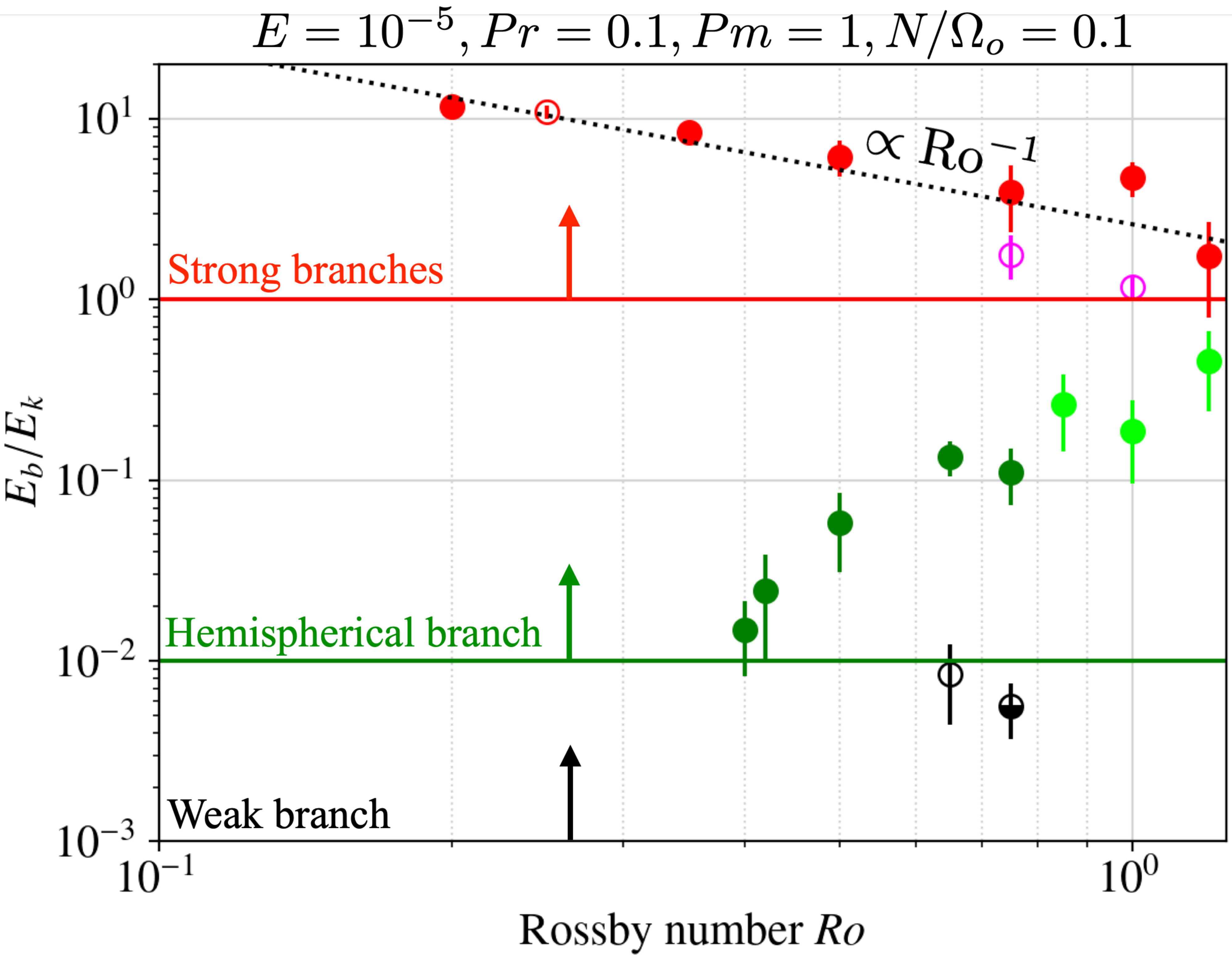}
    \caption{Time-averaged ratio of the magnetic energy to the kinetic energy densities as a function of the Rossby number. The error bars indicate the  standard deviation.}
    \label{fig:energies}
\end{figure}

We find in our set of simulations several dynamo branches represented by different colours in the bifurcation diagram shown in Fig.~\ref{fig:bifurcation}. When the differential rotation is low, the flow can not amplify a weak magnetic field (black crosses), but when $Ro > Ro^c_{\rm W}\sim 0.62$, the magnetic field grows exponentially to reach a metastable or a stable saturated dynamo state (black dots). This kinematic dynamo is driven by an hydrodynamic instability of the Stewartson layer whose threshold is $Ro^c_{\rm hyd}\sim 0.175$ (dashed vertical black line), which is in agreement with~\citet{hollerbach2003}. When $Ro\gtrsim 0.8$, the kinematic growth is followed by a non-linear growth and the system transitions directly to another branch with a larger magnetic energy (green circles). Restarting from a nearby saturated solution or a strong toroidal field with quadrupolar symmetry (mauve dashed arrows), we find that the stability of this branch extends to Rossby number as low as $Ro^c_{\rm H}\sim 0.37 < Ro^c_{\rm W}$, which indicates that this dynamo is subcritical. By starting from a strong toroidal field with dipolar symmetry, we observe that this subcritical branch is in bistability with another one which presents even stronger saturated magnetic fields $B_{\rm rms}\in\left[\SI{4e14}{},\SI{1.1e15}{}\right]\,$G (red circles). This branch is also subcritical since it can be maintained for Rossby numbers as low as $Ro>Ro^{c}_{\rm D}\sim 0.19$. 
Moreover, the two subcritical branches do not only differ by their magnetic field strength but also by their equatorial symmetry, as seen in the 3D snapshots and the surface maps of the magnetic field in Fig.~\ref{fig:bifurcation}. Indeed, the magnetic field shows a dipolar symmetry on the stronger dynamo branch, whereas it is hemispherical on the weaker one. The latter can be interpreted as the superposition of modes with opposite equatorial symmetry~\citep{gallet2009}, which is consistent with the fact that we do find quadrupolar solutions (mauve circles in Fig.~\ref{fig:bifurcation}). These are only metastable for $Ro>Ro^{c}_{\rm Q}\sim 0.7$ and transition to a stable dipolar or hemispherical solution. Finally, we note that the hemispherical dynamos with $Ro\gtrsim0.8$ (light green circles in Figs~\ref{fig:bifurcation} and \ref{fig:energies}) display parity modulations
(i.e. the solution evolves between hemispherical, dipole, and quadrupole symmetric states). This behaviour is reminiscent of the so-called Type~1 modulation identified in other dynamo setups~\citep{knobloch1998,raynaud2016} and likely results from the coupling of modes with opposite parity as the equatorial symmetry breaking of the flow increases at larger Rossby numbers. 

The difference between the three dynamo branches is also clear in Fig.~\ref{fig:energies}, where we see that the hemispherical branch saturates below the equipartition, with an energy ratio increasing with $Ro$ from $\sim\num{0.014}$ up to $\sim\num{0.56}$. By contrast, the dynamos of the dipolar branch are in a super-equipartition state ($E_b/E_k>1$) and follow the magnetostrophic scaling $E_b/E_k\propto Ro^{-1}$ characteristic of the Coriolis-Lorenz force balance \citep{roberts1972,dormy2016,aubert2017,dormy2018,augustson2019,seshasayanan2019,Raynaud2020,schwaiger2019}. This is also confirmed by force balance spectra shown in Fig.~S1 in the Supplemental Materials.

\begin{figure}
    \includegraphics[width=\columnwidth]{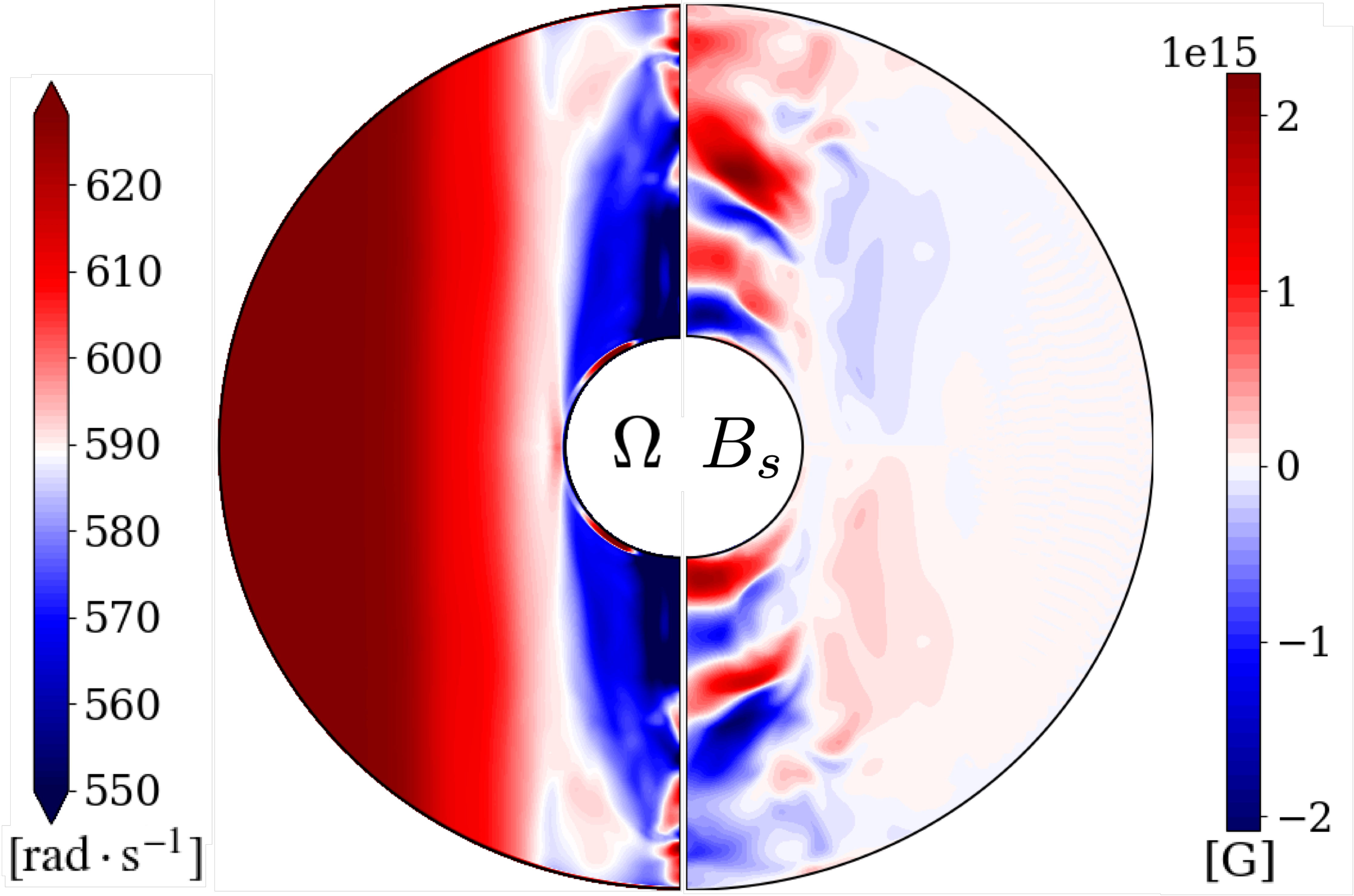}
    \caption{Snapshots of the azimuthal slices of the angular velocity (left) and the magnetic field along the cylindrical radius $s\equiv r\sin{\theta}$ (right) of the dipolar dynamo at $Ro=0.75$.}
    \label{fig:rotBs}
\end{figure}

\begin{figure}
    \includegraphics[width=\columnwidth]{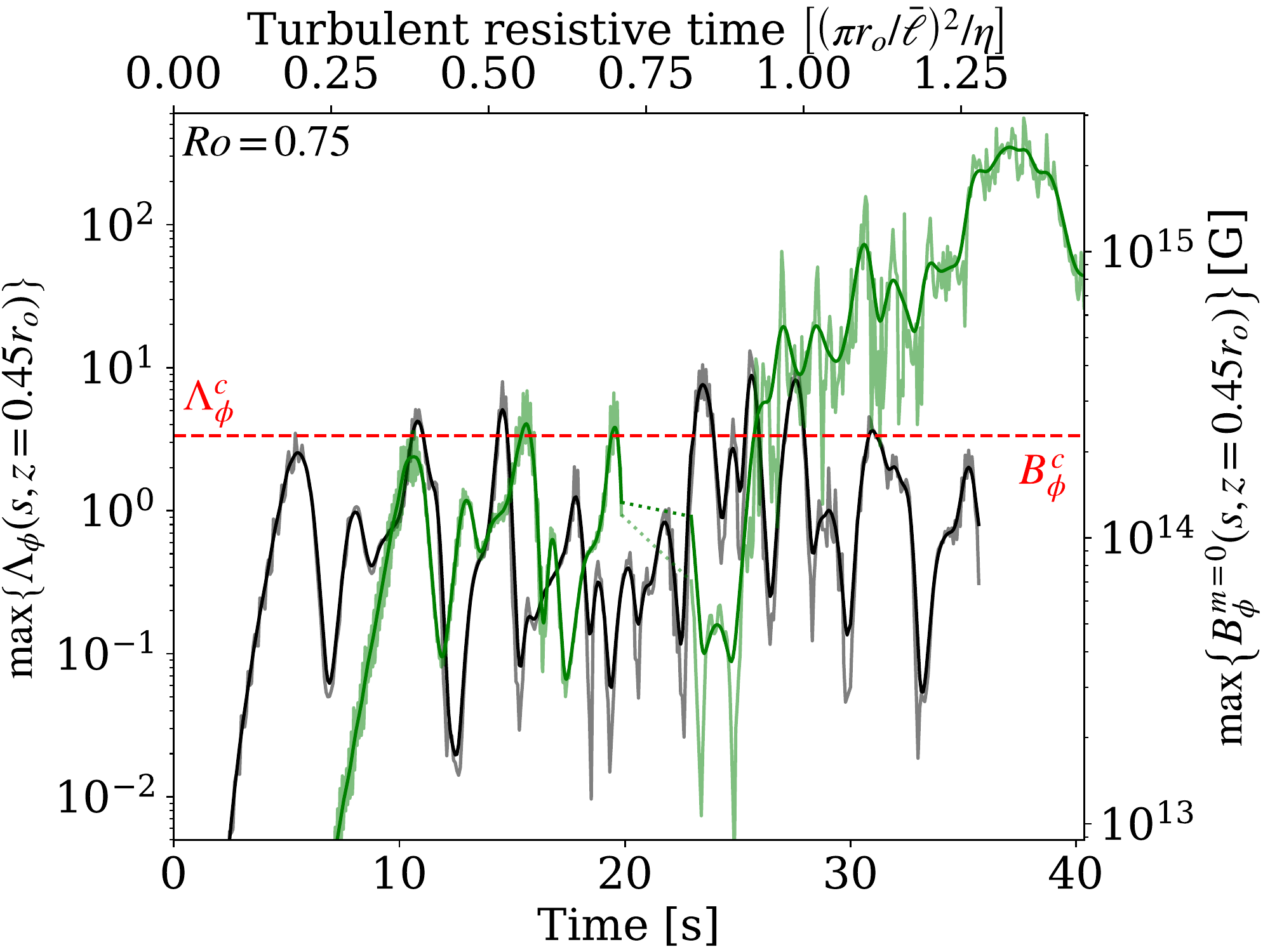}
    \caption{Time series of the maximum along the cylindrical radius $s$ of the axisymmetric toroidal magnetic energy
    measured locally at $z=0.45r_o$,
    for stable (black) and metastable (green) kinematic dynamos
    at $Ro=0.75$. The dashed red line indicates the analytical threshold of the Tayler instability (equation~\ref{eq:threshold}). Dark lines show a running average and dotted green lines around $t\sim\SI{20}{s}$ indicate missing data.}
    \label{fig:BLoc}
\end{figure}

Both subcritical dynamos show magnetic fields concentrated along the rotation axis, which differs significantly from the subcritical solutions found with a negative shear by \citep{petitdemange2023} ; this is also strikingly different from the magnetic field generated on the equatorial plane by the kinematic dynamo (see 3D snapshots of Fig.~\ref{fig:bifurcation}). This suggests that the dipolar and hemispherical dynamos are driven by a different mechanism. We argue that they are driven by the Tayler instability according to the following arguments. First, the axisymmetric toroidal magnetic component is clearly dominant since it contains $53-88\,\%$ of the total magnetic energy. 
Second, the simulations show a poloidal magnetic field with a dominant $m=1$ mode (see Supplemental Materials Figs~S2 and~S3), which is the most unstable mode of the Tayler instability \citep{zahn2007,ma2019}. In the azimuthal cut of the magnetic field component $B_s$ in Fig.~\ref{fig:rotBs},,the Tayler mode also appears clearly close to the poles, where it is expected to develop for a toroidal field generated by the shearing of a poloidal field (see Supplemental Materials Fig.~S4). This is also consistent with the 3D snapshots of the dipolar and hemispherical branches in Fig.~\ref{fig:bifurcation} where the toroidal magnetic field seems prone to a kink instability.  
Third, as in~\citet{petitdemange2023}, the system bifurcates from the kinematic to the hemispherical branch in the vicinity of the threshold of the Tayler instability \citep{spruit1999,spruit2002}
\begin{equation}\label{eq:threshold}
    \Lambda^c_{\phi}\equiv\frac{{B^c_{\phi}}^2}{4\pi\rho\eta\Omega_o}\sim \frac{\chi}{1-\chi}\frac{N}{\Omega_o}\sqrt{\frac{Pr}{E}}\sim 3.3\,.
\end{equation}
Indeed, if we focus on the stable and metastable kinematic solutions found at $Ro=0.75$, we see in Fig.~\ref{fig:BLoc} that the \emph{local} maximum of the toroidal axisymmetric field is in both cases close to the critical value above which it is expected to become unstable. The bifurcation from the kinematic toward the hemispherical branch that is observed for the metastable solution appears hence as the result of turbulent fluctuations departing far enough above the threshold of the Tayler instability.

\begin{figure}
    \includegraphics[width=\columnwidth]{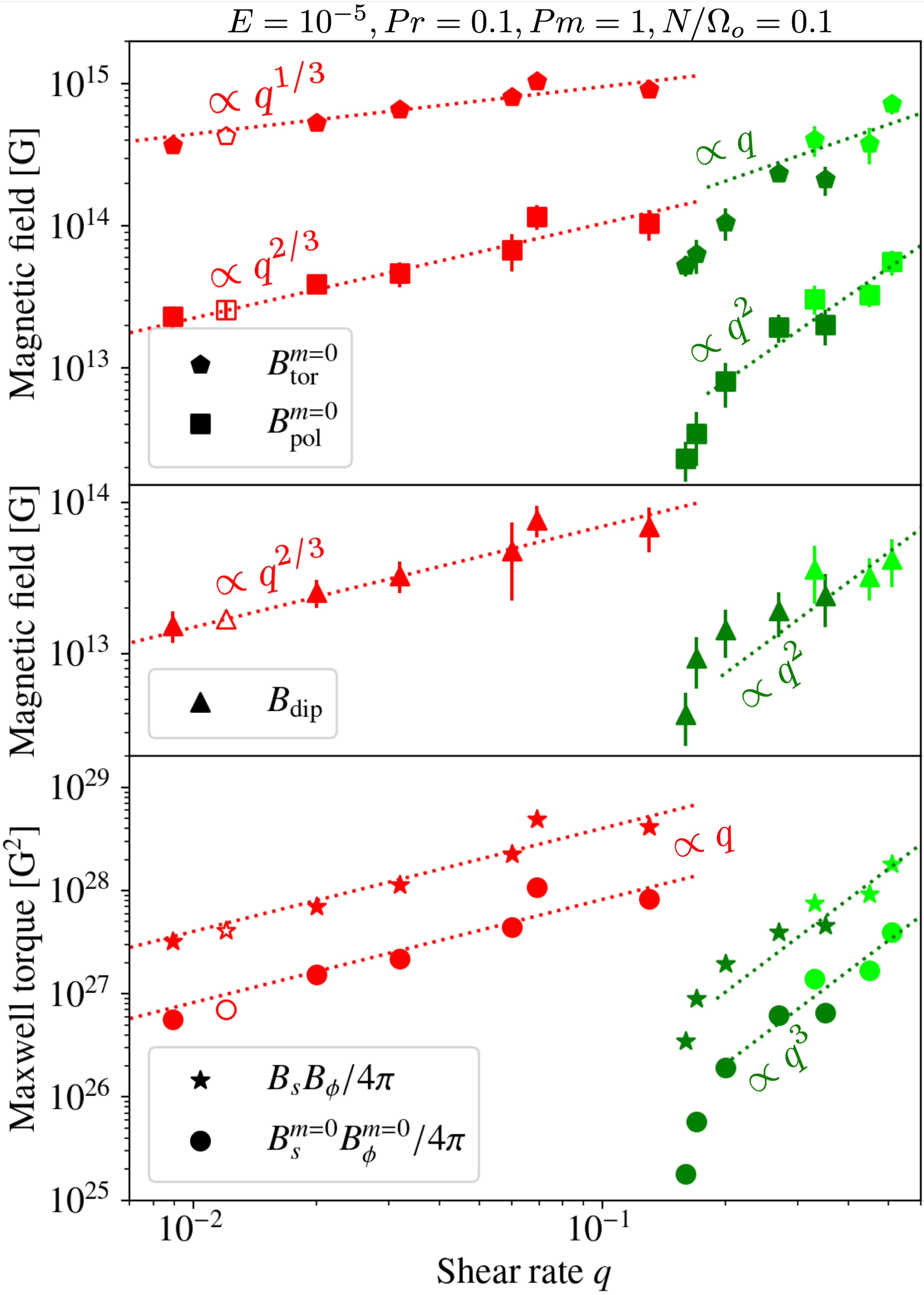}
    \caption{Root mean square (RMS) toroidal and poloidal axisymmetric magnetic fields (top), RMS magnetic dipole (middle), and RMS magnetic torque (bottom) as a function of the time-averaged shear rate measured in the steady state, for the dipolar (red) and hemispherical (green) dynamo branches.
    Dotted lines shows the best fits obtained with Fuller's (red) and Spruit's (green) theoretical scaling laws, respectively.}
    \label{fig:scalings_B}
\end{figure}

Finally, we compare our numerical results to the theoretical predictions regarding the saturation of the Tayler-Spruit dynamo. 
Note that these predictions assume the scale separation $\omega_A \ll \Omega_o \ll N$, where the Alfv\'{e}n frequency is defined by $\omega_A\equiv B_{\phi}/\sqrt{4\pi\rho r_o^2}
\sim 12.1\left(B_{\phi}/10^{15}\,{\rm G}\right)\si{Hz}
$. Our numerical models assume $N/\Omega_o=0.1$ to limit the computational costs, whereas for a typical PNS spun up by fallback to a period of $1-10$~ms we expect $N/\Omega_o\sim 1-10 $. On the other hand, the achieved magnetic field follows the right scale separation with $\omega_A/\Omega_o \lesssim 0.02$, which is expected to determine the saturation mechanism of the Tayler instability \citep{ji2023}.
Figure~\ref{fig:scalings_B} displays the axisymmetric toroidal and poloidal magnetic fields (top), the dipole field (middle) and the Maxwell torque (bottom) as a function of an effective shear rate $q$ measured locally in the saturated state of the dynamo
(see Supplemental Materials Fig.~S5). For the dipolar branch (red), we find that the power laws $B^{m=0}_{\rm tor} \propto q^{0.36\pm0.05}$ 
and $B^{m=0}_{\rm pol} \propto q^{0.62\pm 0.07}$ fit the saturated magnetic field, while we find $B_s B_{\phi} \propto q^{1.0\pm0.02}$ or $B_s^{m=0}B_{\phi}^{m=0}\propto q^{1.1\pm0.04}$, depending on whether we take into account non-axisymmetric contributions to compute the Maxwell torque~$T_{\rm M}$. The scaling exponents are thus in good agreement with the theoretical predictions of~\citet{fuller2019}  $B^{m=0}_{\rm tor}\propto q^{1/3}$, $B^{m=0}_{\rm pol}\propto q^{2/3}$ and $T_{\rm M}\propto q$ (red dotted lines in Fig.~\ref{fig:scalings_B}). Contrary to their prediction, however, our torque is not dominated by the axisymmetric magnetic field, which may be related to their assumption of a stronger stratification. Interestingly, the hemispherical branch (green) does not follow the same scalings: for $q \geq 0.2$, we find $B^{m=0}_{\rm tor}\propto q^{2.1\pm0.31}$ and $B^{m=0}_{\rm pol}\propto q^{2.0\pm 0.28}$ for the magnetic field, and
$B_sB_{\phi}\propto q^{2.7\pm0.40}$ or $B_s^{m=0}B_{\phi}^{m=0} \propto q^{3.8\pm0.70}$ for the Maxwell torque. These results globally support Spruit's predictions~\citep{spruit2002} $B^{m=0}_{\rm tor}\propto q$, $B^{m=0}_{\rm pol}\propto q^{2}$ and $T_{\rm M}\propto q^3$ (green dotted lines)\footnote{In the case of the toroidal magnetic field, the power law index from the fit is in slight tension with the theoretical prediction. However, this tension is not very significant: it is driven mainly by a single data point and disappears if we change the threshold from $q>0.2$ to $q>0.25$ to exclude the model \texttt{Ro0.5as} with $q=0.2$.}. If we focus on the dipole field, we find the following power laws: $B_{\rm dip}\propto q^{0.66\pm0.03}$ and $B_{\rm dip}\propto q^{1.1\pm0.4}$, for the dipolar and hemispherical branches, respectively. The dipole field on the strong branch therefore follows the same scaling as the axisymmetric poloidal field and is only $\sim 33\%$ weaker. 

\section{Conclusions}
\label{sec:discussion}
To conclude, we show that the Tayler-Spruit dynamo also exists in the presence of positive shear. We demonstrate for the first time the existence of two subcritical branches of this dynamo with distinct equatorial symmetries, dipolar and hemispherical. Moreover, the former follows Fuller's theoretical predictions, while the latter is in overall agreement with Spruit's model. 
Compared to the study of~\citet{petitdemange2023} that use a negative shear, our results present a similar dynamical structure, with a bifurcation diagram characterized by a bistability between kinematic and subcritical dynamo solutions. The magnetic field of their Tayler-Spruit dynamo is, however, different since it is characterized by a smaller scale structure localized near the inner boundary in the equatorial plane, and induces a torque scaling according to Spruit's prediction.
Our study shows a magnetic field geometry concentrated near the pole in agreement with the expectation of the Tayler-Spruit dynamo and a more complex physics, with the existence of two different branches that can not be captured by a single scaling law. Extended parameter studies will be needed to further assess the impact of the resistivity and the stratification on this dynamo instability and better constrain its astrophysical implications. 

Our results are of particular importance for stellar evolution models by confirming the existence of the Tayler-Spruit dynamo and by deepening our physical understanding of its complex dynamics. They also give strong support to the new magnetar formation scenario proposed by~\citet{barrere2022}, which relies on the development of a Tayler-instability driven dynamo in the presence of a positive shear. We validate the assumption that the magnetic dipole is a significant fraction of the poloidal magnetic field and follows the same scaling. Extrapolating our results for the dipolar branch to $q\sim1$ as expected in~\citet{barrere2022},
we obtain a magnetic dipole intensity of $\sim\SI{3.2e14}{G}$ and an even stronger axisymmetric toroidal field of $\sim\SI{2.1e15}{G}$.
These orders of magnitude are similar to those found in~\citet{barrere2022} for the same rotation period of $P_o\equiv 2\pi/\Omega_o=\SI{10}{ms}$, and fall right in the magnetar range~\citep{olausen2014}. 

\section*{Acknowledgements}
We thank F. Daniel, C. Gissinger and L. Petitdemange and F. Marcotte for fruitful discussions and for sharing the manuscript \cite{petitdemange2023} before its publication. We also thank A. Igoshev for valuable comments. This work was supported by the European Research Council (MagBURST grant 715368), and the  PNPS and PNHE programs of CNRS/INSU, co-funded by CEA and CNES.
Numerical simulations have been carried out at the CINES on the Jean-Zay supercomputer (DARI project A130410317).

\section*{Data Availability}
The data underlying this article will be shared on reasonable request to the corresponding author.



\bibliographystyle{mnras}
\bibliography{biblio} 

\begin{thebibliography}{}
\makeatletter
\relax
\def\mn@urlcharsother{\let\do\@makeother \do\$\do\&\do\#\do\^\do\_\do\%\do\~}
\def\mn@doi{\begingroup\mn@urlcharsother \@ifnextchar [ {\mn@doi@}
  {\mn@doi@[]}}
\def\mn@doi@[#1]#2{\def\@tempa{#1}\ifx\@tempa\@empty \href
  {http://dx.doi.org/#2} {doi:#2}\else \href {http://dx.doi.org/#2} {#1}\fi
  \endgroup}
\def\mn@eprint#1#2{\mn@eprint@#1:#2::\@nil}
\def\mn@eprint@arXiv#1{\href {http://arxiv.org/abs/#1} {{\tt arXiv:#1}}}
\def\mn@eprint@dblp#1{\href {http://dblp.uni-trier.de/rec/bibtex/#1.xml}
  {dblp:#1}}
\def\mn@eprint@#1:#2:#3:#4\@nil{\def\@tempa {#1}\def\@tempb {#2}\def\@tempc
  {#3}\ifx \@tempc \@empty \let \@tempc \@tempb \let \@tempb \@tempa \fi \ifx
  \@tempb \@empty \def\@tempb {arXiv}\fi \@ifundefined
  {mn@eprint@\@tempb}{\@tempb:\@tempc}{\expandafter \expandafter \csname
  mn@eprint@\@tempb\endcsname \expandafter{\@tempc}}}

\bibitem[\protect\citeauthoryear{{Aubert}, {Gastine}  \& {Fournier}}{{Aubert}
  et~al.}{2017}]{aubert2017}
{Aubert} J.,  {Gastine} T.,   {Fournier} A.,  2017, \mn@doi [Journal of Fluid
  Mechanics] {10.1017/jfm.2016.789}, 813, 558

\bibitem[\protect\citeauthoryear{{Augustson}, {Brun}  \& {Toomre}}{{Augustson}
  et~al.}{2019}]{augustson2019}
{Augustson} K.~C.,  {Brun} A.~S.,   {Toomre} J.,  2019, \mn@doi [Astrophys. J.]
  {10.3847/1538-4357/ab14ea}, 876, 83

\bibitem[\protect\citeauthoryear{{Barr{\`e}re}, {Guilet}, {Reboul-Salze},
  {Raynaud}  \& {Janka}}{{Barr{\`e}re} et~al.}{2022}]{barrere2022}
{Barr{\`e}re} P.,  {Guilet} J.,  {Reboul-Salze} A.,  {Raynaud} R.,   {Janka}
  H.~T.,  2022, \mn@doi [Astron. Astrophys.] {10.1051/0004-6361/202244172},
  668, A79

\bibitem[\protect\citeauthoryear{{Beniamini}, {Hotokezaka}, {van der Horst}  \&
  {Kouveliotou}}{{Beniamini} et~al.}{2019}]{beniamini2019}
{Beniamini} P.,  {Hotokezaka} K.,  {van der Horst} A.,   {Kouveliotou} C.,
  2019, \mn@doi [Mon. Not. Astron. R. Soc.] {10.1093/Mon. Not. Astron. R.
  Soc./stz1391}, 487, 1426

\bibitem[\protect\citeauthoryear{{Bochenek}, {Ravi}, {Belov}, {Hallinan},
  {Kocz}, {Kulkarni}  \& {McKenna}}{{Bochenek} et~al.}{2020}]{bochenek2020}
{Bochenek} C.~D.,  {Ravi} V.,  {Belov} K.~V.,  {Hallinan} G.,  {Kocz} J.,
  {Kulkarni} S.~R.,   {McKenna} D.~L.,  2020, \mn@doi [\nat]
  {10.1038/s41586-020-2872-x}, 587, 59

\bibitem[\protect\citeauthoryear{{Bonanno}, {Guerrero}  \& {Del
  Sordo}}{{Bonanno} et~al.}{2020}]{bonanno2020}
{Bonanno} A.,  {Guerrero} G.,   {Del Sordo} F.,  2020, MEMSAI, 91, 249

\bibitem[\protect\citeauthoryear{{Braithwaite}}{{Braithwaite}}{2006}]{braithwaite2006}
{Braithwaite} J.,  2006, \mn@doi [Astron. Astrophys.]
  {10.1051/0004-6361:20054241}, 449, 451

\bibitem[\protect\citeauthoryear{{Bugli}, {Guilet}, {Obergaulinger},
  {Cerd{\'a}-Dur{\'a}n}  \& {Aloy}}{{Bugli} et~al.}{2020}]{bugli2020}
{Bugli} M.,  {Guilet} J.,  {Obergaulinger} M.,  {Cerd{\'a}-Dur{\'a}n} P.,
  {Aloy} M.~A.,  2020, \mn@doi [Mon. Not. Astron. R. Soc.] {10.1093/Mon. Not.
  Astron. R. Soc./stz3483}, 492, 58

\bibitem[\protect\citeauthoryear{{Bugli}, {Guilet}  \& {Obergaulinger}}{{Bugli}
  et~al.}{2021}]{bugli2021}
{Bugli} M.,  {Guilet} J.,   {Obergaulinger} M.,  2021, \mn@doi [Mon. Not.
  Astron. R. Soc.] {10.1093/Mon. Not. Astron. R. Soc./stab2161}, 507, 443

\bibitem[\protect\citeauthoryear{{Bugli}, {Guilet}, {Foglizzo}  \&
  {Obergaulinger}}{{Bugli} et~al.}{2023}]{bugli2023}
{Bugli} M.,  {Guilet} J.,  {Foglizzo} T.,   {Obergaulinger} M.,  2023, \mn@doi
  [Mon. Not. Astron. R. Soc.] {10.1093/mnras/stad496}, 520, 5622

\bibitem[\protect\citeauthoryear{{Burrows}, {Dessart}, {Livne}, {Ott}  \&
  {Murphy}}{{Burrows} et~al.}{2007}]{burrows2007}
{Burrows} A.,  {Dessart} L.,  {Livne} E.,  {Ott} C.~D.,   {Murphy} J.,  2007,
  \mn@doi [Astrophys. J.] {10.1086/519161}, 664, 416

\bibitem[\protect\citeauthoryear{{CHIME/FRB Collaboration} et~al.,}{{CHIME/FRB
  Collaboration} et~al.}{2020}]{chime2020}
{CHIME/FRB Collaboration} et~al., 2020, \mn@doi [\nat]
  {10.1038/s41586-020-2863-y}, 587, 54

\bibitem[\protect\citeauthoryear{{Cantiello}, {Mankovich}, {Bildsten},
  {Christensen-Dalsgaard}  \& {Paxton}}{{Cantiello}
  et~al.}{2014}]{cantiello2014}
{Cantiello} M.,  {Mankovich} C.,  {Bildsten} L.,  {Christensen-Dalsgaard} J.,
  {Paxton} B.,  2014, \mn@doi [Astrophys. J.] {10.1088/0004-637X/788/1/93},
  788, 93

\bibitem[\protect\citeauthoryear{{Chan}, {M{\"u}ller}  \& {Heger}}{{Chan}
  et~al.}{2020}]{chan2020}
{Chan} C.,  {M{\"u}ller} B.,   {Heger} A.,  2020, \mn@doi [Mon. Not. Astron. R.
  Soc.] {10.1093/Mon. Not. Astron. R. Soc./staa1431}, 495, 3751

\bibitem[\protect\citeauthoryear{{Coti Zelati}, {Rea}, {Pons}, {Campana}  \&
  {Esposito}}{{Coti Zelati} et~al.}{2018}]{coti2018}
{Coti Zelati} F.,  {Rea} N.,  {Pons} J.~A.,  {Campana} S.,   {Esposito} P.,
  2018, \mn@doi [Mon. Not. Astron. R. Soc.] {10.1093/Mon. Not. Astron. R.
  Soc./stx2679}, 474, 961

\bibitem[\protect\citeauthoryear{{Coti Zelati} et~al.,}{{Coti Zelati}
  et~al.}{2021}]{coti2021}
{Coti Zelati} F.,  et~al., 2021, \mn@doi [Astrophys. J. Lett.]
  {10.3847/2041-8213/abda52}, 907, L34

\bibitem[\protect\citeauthoryear{{Denissenkov} \& {Pinsonneault}}{{Denissenkov}
  \& {Pinsonneault}}{2007}]{denissenkov2007}
{Denissenkov} P.~A.,  {Pinsonneault} M.,  2007, \mn@doi [Astrophys. J.]
  {10.1086/510345}, 655, 1157

\bibitem[\protect\citeauthoryear{{Dessart}, {Burrows}, {Livne}  \&
  {Ott}}{{Dessart} et~al.}{2008}]{dessart2008}
{Dessart} L.,  {Burrows} A.,  {Livne} E.,   {Ott} C.~D.,  2008, \mn@doi
  [Astrophys. J. Lett.] {10.1086/527519}, 673, L43

\bibitem[\protect\citeauthoryear{{Dormy}}{{Dormy}}{2016}]{dormy2016}
{Dormy} E.,  2016, \mn@doi [J. Fluid Mech.] {10.1017/jfm.2015.747}, 789, 500

\bibitem[\protect\citeauthoryear{{Dormy}, {Oruba}  \& {Petitdemange}}{{Dormy}
  et~al.}{2018}]{dormy2018}
{Dormy} E.,  {Oruba} L.,   {Petitdemange} L.,  2018, \mn@doi [Fluid Dyn. Res.]
  {10.1088/1873-7005/aa769c}, 50, 011415

\bibitem[\protect\citeauthoryear{{Eggenberger}, {Maeder}  \&
  {Meynet}}{{Eggenberger} et~al.}{2005}]{eggenberger2005}
{Eggenberger} P.,  {Maeder} A.,   {Meynet} G.,  2005, \mn@doi [Astron.
  Astrophys.] {10.1051/0004-6361:200500156}, 440, L9

\bibitem[\protect\citeauthoryear{{Eggenberger}, {Buldgen}  \&
  {Salmon}}{{Eggenberger} et~al.}{2019a}]{eggenberger2019b}
{Eggenberger} P.,  {Buldgen} G.,   {Salmon} S.~J.~A.~J.,  2019a, \mn@doi
  [Astron. Astrophys.] {10.1051/0004-6361/201935509}, 626, L1

\bibitem[\protect\citeauthoryear{{Eggenberger}, {den Hartogh}, {Buldgen},
  {Meynet}, {Salmon}  \& {Deheuvels}}{{Eggenberger}
  et~al.}{2019b}]{eggenberger2019a}
{Eggenberger} P.,  {den Hartogh} J.~W.,  {Buldgen} G.,  {Meynet} G.,  {Salmon}
  S.~J.~A.~J.,   {Deheuvels} S.,  2019b, \mn@doi [Astron. Astrophys.]
  {10.1051/0004-6361/201936348}, 631, L6

\bibitem[\protect\citeauthoryear{{Evans} et~al.,}{{Evans}
  et~al.}{1980}]{evans1980}
{Evans} W.~D.,  et~al., 1980, \mn@doi [Astrophys. J. Lett.] {10.1086/183222},
  237, L7

\bibitem[\protect\citeauthoryear{{Ferrario} \& {Wickramasinghe}}{{Ferrario} \&
  {Wickramasinghe}}{2006}]{ferrario2006}
{Ferrario} L.,  {Wickramasinghe} D.,  2006, \mn@doi [Mon. Not. Astron. R. Soc.]
  {10.1111/j.1365-2966.2006.10058.x}, 367, 1323

\bibitem[\protect\citeauthoryear{{Fuller}, {Piro}  \& {Jermyn}}{{Fuller}
  et~al.}{2019}]{fuller2019}
{Fuller} J.,  {Piro} A.~L.,   {Jermyn} A.~S.,  2019, \mn@doi [Mon. Not. Astron.
  R. Soc.] {10.1093/Mon. Not. Astron. R. Soc./stz514}, 485, 3661

\bibitem[\protect\citeauthoryear{{Gallet} \& {P{\'e}tr{\'e}lis}}{{Gallet} \&
  {P{\'e}tr{\'e}lis}}{2009}]{gallet2009}
{Gallet} B.,  {P{\'e}tr{\'e}lis} F.,  2009, \mn@doi [\pre]
  {10.1103/PhysRevE.80.035302}, 80, 035302

\bibitem[\protect\citeauthoryear{{Gastine} \& {Wicht}}{{Gastine} \&
  {Wicht}}{2012}]{gastine2012}
{Gastine} T.,  {Wicht} J.,  2012, \mn@doi [Icarus]
  {10.1016/j.icarus.2012.03.018}, 219, 428

\bibitem[\protect\citeauthoryear{{Gill} \& {Heyl}}{{Gill} \&
  {Heyl}}{2007}]{gill2007}
{Gill} R.,  {Heyl} J.,  2007, \mn@doi [Mon. Not. Astron. R. Soc.]
  {10.1111/j.1365-2966.2007.12254.x}, 381, 52

\bibitem[\protect\citeauthoryear{{Gompertz}, {O'Brien}  \& {Wynn}}{{Gompertz}
  et~al.}{2014}]{gompertz2014}
{Gompertz} B.~P.,  {O'Brien} P.~T.,   {Wynn} G.~A.,  2014, \mn@doi [Mon. Not.
  Astron. R. Soc.] {10.1093/Mon. Not. Astron. R. Soc./stt2165}, 438, 240

\bibitem[\protect\citeauthoryear{{Gotz} et~al.,}{{Gotz}
  et~al.}{2006}]{gotz2006}
{Gotz} D.,  et~al., 2006, The Astronomer's Telegram, 953, 1

\bibitem[\protect\citeauthoryear{{Griffiths}, {Eggenberger}, {Meynet}, {Moyano}
   \& {Aloy}}{{Griffiths} et~al.}{2022}]{griffiths2022}
{Griffiths} A.,  {Eggenberger} P.,  {Meynet} G.,  {Moyano} F.,   {Aloy}
  M.-{\'A}.,  2022, \mn@doi [Astron. Astrophys.] {10.1051/0004-6361/202243599},
  665, A147

\bibitem[\protect\citeauthoryear{{Gubbins} \& {Zhang}}{{Gubbins} \&
  {Zhang}}{1993}]{gubbins1993}
{Gubbins} D.,  {Zhang} K.,  1993, \mn@doi [Physics of the Earth and Planetary
  Interiors] {10.1016/0031-9201(93)90003-R}, \href
  {https://ui.adsabs.harvard.edu/abs/1993PEPI...75..225G} {75, 225}

\bibitem[\protect\citeauthoryear{{Guilet}, {Reboul-Salze}, {Raynaud}, {Bugli}
  \& {Gallet}}{{Guilet} et~al.}{2022}]{guilet2022}
{Guilet} J.,  {Reboul-Salze} A.,  {Raynaud} R.,  {Bugli} M.,   {Gallet} B.,
  2022, \mn@doi [Mon. Not. Astron. R. Soc.] {10.1093/Mon. Not. Astron. R.
  Soc./stac2499}, 516, 4346

\bibitem[\protect\citeauthoryear{{Hollerbach}}{{Hollerbach}}{2003}]{hollerbach2003}
{Hollerbach} R.,  2003, \mn@doi [J. Fluid Mech.] {10.1017/S0022112003005676},
  492, 289

\bibitem[\protect\citeauthoryear{{Hu} \& {Lou}}{{Hu} \& {Lou}}{2009}]{hu2009}
{Hu} R.-Y.,  {Lou} Y.-Q.,  2009, \mn@doi [Mon. Not. Astron. R. Soc.]
  {10.1111/j.1365-2966.2009.14648.x}, 396, 878

\bibitem[\protect\citeauthoryear{{Hurley} et~al.,}{{Hurley}
  et~al.}{1999}]{hurley1999}
{Hurley} K.,  et~al., 1999, \mn@doi [\nat] {10.1038/16199}, 397, 41

\bibitem[\protect\citeauthoryear{{Hurley} et~al.,}{{Hurley}
  et~al.}{2005}]{hurley2005}
{Hurley} K.,  et~al., 2005, \mn@doi [\nat] {10.1038/nature03519}, 434, 1098

\bibitem[\protect\citeauthoryear{{Janka}, {Wongwathanarat}  \&
  {Kramer}}{{Janka} et~al.}{2022}]{janka2021}
{Janka} H.-T.,  {Wongwathanarat} A.,   {Kramer} M.,  2022, \mn@doi [Astrophys.
  J.] {10.3847/1538-4357/ac403c}, 926, 9

\bibitem[\protect\citeauthoryear{{Ji}, {Fuller}  \& {Lecoanet}}{{Ji}
  et~al.}{2023}]{ji2023}
{Ji} S.,  {Fuller} J.,   {Lecoanet} D.,  2023, \mn@doi [Mon. Not. Astron. R.
  Soc.] {10.1093/mnras/stad910}, 521, 5372

\bibitem[\protect\citeauthoryear{{Knobloch}, {Tobias}  \& {Weiss}}{{Knobloch}
  et~al.}{1998}]{knobloch1998}
{Knobloch} E.,  {Tobias} S.~M.,   {Weiss} N.~O.,  1998, \mn@doi [Mon. Not.
  Astron. R. Soc.] {10.1046/j.1365-8711.1998.01572.x}, 297, 1123

\bibitem[\protect\citeauthoryear{{Kouveliotou} et~al.,}{{Kouveliotou}
  et~al.}{1994}]{kouveliotou1994}
{Kouveliotou} C.,  et~al., 1994, \mn@doi [\nat] {10.1038/368125a0}, 368, 125

\bibitem[\protect\citeauthoryear{{Kuroda}, {Arcones}, {Takiwaki}  \&
  {Kotake}}{{Kuroda} et~al.}{2020}]{kuroda2020}
{Kuroda} T.,  {Arcones} A.,  {Takiwaki} T.,   {Kotake} K.,  2020, \mn@doi
  [Astrophys. J.] {10.3847/1538-4357/ab9308}, 896, 102

\bibitem[\protect\citeauthoryear{{L{\"u}} \& {Zhang}}{{L{\"u}} \&
  {Zhang}}{2014}]{lu2014}
{L{\"u}} H.-J.,  {Zhang} B.,  2014, \mn@doi [Astrophys. J.]
  {10.1088/0004-637X/785/1/74}, 785, 74

\bibitem[\protect\citeauthoryear{{Ma} \& {Fuller}}{{Ma} \&
  {Fuller}}{2019}]{ma2019}
{Ma} L.,  {Fuller} J.,  2019, \mn@doi [Mon. Not. Astron. R. Soc.] {10.1093/Mon.
  Not. Astron. R. Soc./stz2009}, 488, 4338

\bibitem[\protect\citeauthoryear{{Martin}, {Rea}, {Torres}  \&
  {Papitto}}{{Martin} et~al.}{2014}]{martin2014}
{Martin} J.,  {Rea} N.,  {Torres} D.~F.,   {Papitto} A.,  2014, \mn@doi [Mon.
  Not. Astron. R. Soc.] {10.1093/Mon. Not. Astron. R. Soc./stu1594}, 444, 2910

\bibitem[\protect\citeauthoryear{{Masada}, {Takiwaki}  \& {Kotake}}{{Masada}
  et~al.}{2022}]{masada2022}
{Masada} Y.,  {Takiwaki} T.,   {Kotake} K.,  2022, \mn@doi [Astrophys. J.]
  {10.3847/1538-4357/ac34f6}, 924, 75

\bibitem[\protect\citeauthoryear{{Metzger}, {Quataert}  \&
  {Thompson}}{{Metzger} et~al.}{2008}]{metzger2008}
{Metzger} B.~D.,  {Quataert} E.,   {Thompson} T.~A.,  2008, \mn@doi [Mon. Not.
  Astron. R. Soc.] {10.1111/j.1365-2966.2008.12923.x}, 385, 1455

\bibitem[\protect\citeauthoryear{{M{\"o}sta} et~al.,}{{M{\"o}sta}
  et~al.}{2014}]{moesta2014}
{M{\"o}sta} P.,  et~al., 2014, \mn@doi [Astrophys. J. Lett.]
  {10.1088/2041-8205/785/2/L29}, 785, L29

\bibitem[\protect\citeauthoryear{{Obergaulinger} \& {Aloy}}{{Obergaulinger} \&
  {Aloy}}{2020}]{Obergaulinger2020}
{Obergaulinger} M.,  {Aloy} M.~{\'A}.,  2020, \mn@doi [Mon. Not. Astron. R.
  Soc.] {10.1093/Mon. Not. Astron. R. Soc./staa096}, 492, 4613

\bibitem[\protect\citeauthoryear{{Obergaulinger} \& {Aloy}}{{Obergaulinger} \&
  {Aloy}}{2021}]{Obergaulinger2021}
{Obergaulinger} M.,  {Aloy} M.~{\'A}.,  2021, \mn@doi [Mon. Not. Astron. R.
  Soc.] {10.1093/Mon. Not. Astron. R. Soc./stab295}, 503, 4942

\bibitem[\protect\citeauthoryear{{Obergaulinger} \& {Aloy}}{{Obergaulinger} \&
  {Aloy}}{2022}]{Obergaulinger2022}
{Obergaulinger} M.,  {Aloy} M.~{\'A}.,  2022, \mn@doi [Mon. Not. Astron. R.
  Soc.] {10.1093/Mon. Not. Astron. R. Soc./stac613}, 512, 2489

\bibitem[\protect\citeauthoryear{{Obergaulinger}, {Cerd{\'a}-Dur{\'a}n},
  {M{\"u}ller}  \& {Aloy}}{{Obergaulinger} et~al.}{2009}]{obergaulinger2009}
{Obergaulinger} M.,  {Cerd{\'a}-Dur{\'a}n} P.,  {M{\"u}ller} E.,   {Aloy}
  M.~A.,  2009, \mn@doi [Astron. Astrophys.] {10.1051/0004-6361/200811323},
  498, 241

\bibitem[\protect\citeauthoryear{{Olausen} \& {Kaspi}}{{Olausen} \&
  {Kaspi}}{2014}]{olausen2014}
{Olausen} S.~A.,  {Kaspi} V.~M.,  2014, \mn@doi [Astrophys. J.s]
  {10.1088/0067-0049/212/1/6}, 212, 6

\bibitem[\protect\citeauthoryear{Petitdemange, Marcotte  \&
  Gissinger}{Petitdemange et~al.}{2023}]{petitdemange2023}
Petitdemange L.,  Marcotte F.,   Gissinger C.,  2023, \mn@doi [Science]
  {10.1126/science.abk2169}, 379, 300

\bibitem[\protect\citeauthoryear{{Pitts} \& {Tayler}}{{Pitts} \&
  {Tayler}}{1985}]{pitts1985}
{Pitts} E.,  {Tayler} R.~J.,  1985, \mn@doi [Mon. Not. Astron. R. Soc.]
  {10.1093/Mon. Not. Astron. R. Soc./216.2.139}, 216, 139

\bibitem[\protect\citeauthoryear{{Raynaud} \& {Tobias}}{{Raynaud} \&
  {Tobias}}{2016}]{raynaud2016}
{Raynaud} R.,  {Tobias} S.~M.,  2016, \mn@doi [J. Fluid Mech.]
  {10.1017/jfm.2016.407}, 799, R6

\bibitem[\protect\citeauthoryear{{Raynaud}, {Guilet}, {Janka}  \&
  {Gastine}}{{Raynaud} et~al.}{2020}]{Raynaud2020}
{Raynaud} R.,  {Guilet} J.,  {Janka} H.-T.,   {Gastine} T.,  2020, \mn@doi
  [Sci. Adv.] {10.1126/sciadv.aay2732}, 6, eaay2732

\bibitem[\protect\citeauthoryear{{Raynaud}, {Cerd{\'a}-Dur{\'a}n}  \&
  {Guilet}}{{Raynaud} et~al.}{2022}]{raynaud2022}
{Raynaud} R.,  {Cerd{\'a}-Dur{\'a}n} P.,   {Guilet} J.,  2022, \mn@doi [Mon.
  Not. Astron. R. Soc.] {10.1093/Mon. Not. Astron. R. Soc./stab3109}, 509, 3410

\bibitem[\protect\citeauthoryear{{Reboul-Salze}, {Guilet}, {Raynaud}  \&
  {Bugli}}{{Reboul-Salze} et~al.}{2021}]{reboul2021a}
{Reboul-Salze} A.,  {Guilet} J.,  {Raynaud} R.,   {Bugli} M.,  2021, \mn@doi
  [Astron. Astrophys.] {10.1051/0004-6361/202038369}, 645, A109

\bibitem[\protect\citeauthoryear{{Reboul-Salze}, {Guilet}, {Raynaud}  \&
  {Bugli}}{{Reboul-Salze} et~al.}{2022}]{reboul2022}
{Reboul-Salze} A.,  {Guilet} J.,  {Raynaud} R.,   {Bugli} M.,  2022, \mn@doi
  [Astron. Astrophys.] {10.1051/0004-6361/202142368}, 667, A94

\bibitem[\protect\citeauthoryear{{Roberts} \& {Soward}}{{Roberts} \&
  {Soward}}{1972}]{roberts1972}
{Roberts} P.~H.,  {Soward} A.~M.,  1972, \mn@doi [Annual Review of Fluid
  Mechanics] {10.1146/annurev.fl.04.010172.001001}, 4, 117

\bibitem[\protect\citeauthoryear{{Schaeffer}}{{Schaeffer}}{2013}]{schaeffer2013}
{Schaeffer} N.,  2013, \mn@doi [Geochemistry, Geophysics, Geosystems]
  {10.1002/ggge.20071}, 14, 751

\bibitem[\protect\citeauthoryear{{Schneider}, {Ohlmann}, {Podsiadlowski},
  {R{\"o}pke}, {Balbus}  \& {Pakmor}}{{Schneider} et~al.}{2020}]{schneider2020}
{Schneider} F.~R.~N.,  {Ohlmann} S.~T.,  {Podsiadlowski} P.,  {R{\"o}pke}
  F.~K.,  {Balbus} S.~A.,   {Pakmor} R.,  2020, \mn@doi [Mon. Not. Astron. R.
  Soc.] {10.1093/Mon. Not. Astron. R. Soc./staa1326}, 495, 2796

\bibitem[\protect\citeauthoryear{{Schwaiger}, {Gastine}  \&
  {Aubert}}{{Schwaiger} et~al.}{2019}]{schwaiger2019}
{Schwaiger} T.,  {Gastine} T.,   {Aubert} J.,  2019, \mn@doi [Geophysical
  Journal International] {10.1093/gji/ggz192}, 219, S101

\bibitem[\protect\citeauthoryear{{Seshasayanan} \& {Gallet}}{{Seshasayanan} \&
  {Gallet}}{2019}]{seshasayanan2019}
{Seshasayanan} K.,  {Gallet} B.,  2019, \mn@doi [J. Fluid Mech.]
  {10.1017/jfm.2019.12}, 864, 971

\bibitem[\protect\citeauthoryear{{Skoutnev}, {Squire}  \&
  {Bhattacharjee}}{{Skoutnev} et~al.}{2022}]{skoutnev2022}
{Skoutnev} V.,  {Squire} J.,   {Bhattacharjee} A.,  2022, \mn@doi [Mon. Not.
  Astron. R. Soc.] {10.1093/Mon. Not. Astron. R. Soc./stac2676}, 517, 526

\bibitem[\protect\citeauthoryear{{Spruit}}{{Spruit}}{1999}]{spruit1999}
{Spruit} H.~C.,  1999, Astron. Astrophys., 349, 189

\bibitem[\protect\citeauthoryear{{Spruit}}{{Spruit}}{2002}]{spruit2002}
{Spruit} H.~C.,  2002, \mn@doi [Astron. Astrophys.]
  {10.1051/0004-6361:20011465}, 381, 923

\bibitem[\protect\citeauthoryear{{Svinkin} et~al.,}{{Svinkin}
  et~al.}{2021}]{svinkin2021}
{Svinkin} D.,  et~al., 2021, \mn@doi [\nat] {10.1038/s41586-020-03076-9}, 589,
  211

\bibitem[\protect\citeauthoryear{{Takiwaki}, {Kotake}  \& {Sato}}{{Takiwaki}
  et~al.}{2009}]{takiwaki2009}
{Takiwaki} T.,  {Kotake} K.,   {Sato} K.,  2009, \mn@doi [Astrophys. J.]
  {10.1088/0004-637X/691/2/1360}, 691, 1360

\bibitem[\protect\citeauthoryear{{Tayler}}{{Tayler}}{1973}]{tayler1973}
{Tayler} R.~J.,  1973, \mn@doi [Mon. Not. Astron. R. Soc.] {10.1093/Mon. Not.
  Astron. R. Soc./161.4.365}, 161, 365

\bibitem[\protect\citeauthoryear{{Thompson} \& {Duncan}}{{Thompson} \&
  {Duncan}}{1993}]{thompson1993}
{Thompson} C.,  {Duncan} R.~C.,  1993, \mn@doi [Astrophys. J.]
  {10.1086/172580}, 408, 194

\bibitem[\protect\citeauthoryear{{Vink} \& {Kuiper}}{{Vink} \&
  {Kuiper}}{2006}]{vink2006}
{Vink} J.,  {Kuiper} L.,  2006, \mn@doi [Mon. Not. Astron. R. Soc.]
  {10.1111/j.1745-3933.2006.00178.x}, 370, L14

\bibitem[\protect\citeauthoryear{{White}, {Burrows}, {Coleman}  \&
  {Vartanyan}}{{White} et~al.}{2022}]{white2022}
{White} C.~J.,  {Burrows} A.,  {Coleman} M. S.~B.,   {Vartanyan} D.,  2022,
  \mn@doi [Astrophys. J.] {10.3847/1538-4357/ac4507}, 926, 111

\bibitem[\protect\citeauthoryear{{Wicht}}{{Wicht}}{2002}]{wicht2002}
{Wicht} J.,  2002, \mn@doi [Physics of the Earth and Planetary Interiors]
  {10.1016/S0031-9201(02)00078-X}, 132, 281

\bibitem[\protect\citeauthoryear{{Woods} \& {Thompson}}{{Woods} \&
  {Thompson}}{2006}]{woods2006}
{Woods} P.~M.,  {Thompson} C.,  2006, in , Vol.~39, Compact stellar X-ray
  sources.
pp 547--586, \mn@doi{10.48550/arXiv.astro-ph/0406133}

\bibitem[\protect\citeauthoryear{{Zahn}, {Brun}  \& {Mathis}}{{Zahn}
  et~al.}{2007}]{zahn2007}
{Zahn} J.~P.,  {Brun} A.~S.,   {Mathis} S.,  2007, \mn@doi [Astron. Astrophys.]
  {10.1051/0004-6361:20077653}, 474, 145

\bibitem[\protect\citeauthoryear{{Zhou}, {Vink}, {Safi-Harb}  \&
  {Miceli}}{{Zhou} et~al.}{2019}]{zhou2019}
{Zhou} P.,  {Vink} J.,  {Safi-Harb} S.,   {Miceli} M.,  2019, \mn@doi [Astron.
  Astrophys.] {10.1051/0004-6361/201936002}, 629, A51

\bibitem[\protect\citeauthoryear{{den Hartogh}, {Eggenberger}  \&
  {Deheuvels}}{{den Hartogh} et~al.}{2020}]{denhartog2020}
{den Hartogh} J.~W.,  {Eggenberger} P.,   {Deheuvels} S.,  2020, \mn@doi
  [Astron. Astrophys.] {10.1051/0004-6361/202037568}, 634, L16

\makeatother
\end{thebibliography}


\begin{thebibliography}{}
\makeatletter
\relax
\def\mn@urlcharsother{\let\do\@makeother \do\$\do\&\do\#\do\^\do\_\do\%\do\~}
\def\mn@doi{\begingroup\mn@urlcharsother \@ifnextchar [ {\mn@doi@}
  {\mn@doi@[]}}
\def\mn@doi@[#1]#2{\def\@tempa{#1}\ifx\@tempa\@empty \href
  {http://dx.doi.org/#2} {doi:#2}\else \href {http://dx.doi.org/#2} {#1}\fi
  \endgroup}
\def\mn@eprint#1#2{\mn@eprint@#1:#2::\@nil}
\def\mn@eprint@arXiv#1{\href {http://arxiv.org/abs/#1} {{\tt arXiv:#1}}}
\def\mn@eprint@dblp#1{\href {http://dblp.uni-trier.de/rec/bibtex/#1.xml}
  {dblp:#1}}
\def\mn@eprint@#1:#2:#3:#4\@nil{\def\@tempa {#1}\def\@tempb {#2}\def\@tempc
  {#3}\ifx \@tempc \@empty \let \@tempc \@tempb \let \@tempb \@tempa \fi \ifx
  \@tempb \@empty \def\@tempb {arXiv}\fi \@ifundefined
  {mn@eprint@\@tempb}{\@tempb:\@tempc}{\expandafter \expandafter \csname
  mn@eprint@\@tempb\endcsname \expandafter{\@tempc}}}

\bibitem[\protect\citeauthoryear{{Aubert}, {Gastine}  \& {Fournier}}{{Aubert}
  et~al.}{2017}]{aubert2017}
{Aubert} J.,  {Gastine} T.,   {Fournier} A.,  2017, \mn@doi [Journal of Fluid
  Mechanics] {10.1017/jfm.2016.789}, 813, 558

\bibitem[\protect\citeauthoryear{{Barr{\`e}re}, {Guilet}, {Reboul-Salze},
  {Raynaud}  \& {Janka}}{{Barr{\`e}re} et~al.}{2022}]{barrere2022}
{Barr{\`e}re} P.,  {Guilet} J.,  {Reboul-Salze} A.,  {Raynaud} R.,   {Janka}
  H.~T.,  2022, \mn@doi [Astron. Astrophys.] {10.1051/0004-6361/202244172},
  668, A79

\bibitem[\protect\citeauthoryear{{Cutler} \& {Lindblom}}{{Cutler} \&
  {Lindblom}}{1987}]{cutler1987}
{Cutler} C.,  {Lindblom} L.,  1987, \mn@doi [Astrophys. J.] {10.1086/165052},
  314, 234

\bibitem[\protect\citeauthoryear{{Gastine} \& {Wicht}}{{Gastine} \&
  {Wicht}}{2012}]{gastine2012}
{Gastine} T.,  {Wicht} J.,  2012, \mn@doi [Icarus]
  {10.1016/j.icarus.2012.03.018}, 219, 428

\bibitem[\protect\citeauthoryear{{Goossens} \& {Tayler}}{{Goossens} \&
  {Tayler}}{1980}]{goossens1980b}
{Goossens} M.,  {Tayler} R.~J.,  1980, \mn@doi [\mnras]
  {10.1093/mnras/193.4.833}, 193, 833

\bibitem[\protect\citeauthoryear{{Guilet}, {M{\"u}ller}  \& {Janka}}{{Guilet}
  et~al.}{2015}]{guilet2015b}
{Guilet} J.,  {M{\"u}ller} E.,   {Janka} H.-T.,  2015, \mn@doi [\mnras]
  {10.1093/mnras/stu2550}, 447, 3992

\bibitem[\protect\citeauthoryear{{H{\"u}depohl}}{{H{\"u}depohl}}{2014}]{hudepohl2014}
{H{\"u}depohl} L.,  2014, PhD thesis, Technical University of Munich, Germany

\bibitem[\protect\citeauthoryear{{Kelly}}{{Kelly}}{1973}]{kelly1973}
{Kelly} D.~C.,  1973, \mn@doi [\apj] {10.1086/151898}, \href
  {https://ui.adsabs.harvard.edu/abs/1973ApJ...179..599K} {179, 599}

\bibitem[\protect\citeauthoryear{Lee}{Lee}{1950}]{lee1950}
Lee T.~D.,  1950, \mn@doi [The Astrophysical Journal] {10.1086/145306}, 111,
  625

\bibitem[\protect\citeauthoryear{{Schaeffer}}{{Schaeffer}}{2013}]{schaeffer2013}
{Schaeffer} N.,  2013, \mn@doi [Geochemistry, Geophysics, Geosystems]
  {10.1002/ggge.20071}, 14, 751

\bibitem[\protect\citeauthoryear{{Schwaiger}, {Gastine}  \&
  {Aubert}}{{Schwaiger} et~al.}{2019}]{schwaiger2019}
{Schwaiger} T.,  {Gastine} T.,   {Aubert} J.,  2019, \mn@doi [Geophysical
  Journal International] {10.1093/gji/ggz192}, 219, S101

\bibitem[\protect\citeauthoryear{{Tayler}}{{Tayler}}{1973}]{tayler1973}
{Tayler} R.~J.,  1973, \mn@doi [Mon. Not. Astron. R. Soc.] {10.1093/Mon. Not.
  Astron. R. Soc./161.4.365}, 161, 365

\bibitem[\protect\citeauthoryear{{Thompson} \& {Duncan}}{{Thompson} \&
  {Duncan}}{1993}]{thompson1993}
{Thompson} C.,  {Duncan} R.~C.,  1993, \mn@doi [Astrophys. J.]
  {10.1086/172580}, 408, 194

\bibitem[\protect\citeauthoryear{{Wicht}}{{Wicht}}{2002}]{wicht2002}
{Wicht} J.,  2002, \mn@doi [Physics of the Earth and Planetary Interiors]
  {10.1016/S0031-9201(02)00078-X}, 132, 281

\makeatother
\end{thebibliography}



\bsp	
\label{lastpage}
\end{document}


\label{firstpage}
\pagerange{\pageref{firstpage}--\pageref{lastpage}}
\maketitle

In these Supplemental Materials, we present in more details our setup and provide further analyses of the dynamo simulations (force balance, time-averaged spectra, instability criterion, shear rate).

\section{Set up}

\subsection{Governing equations}\label{sec:equations}
We model the proto-neutron star differential rotation as a spherical, stably stratified Couette flow. In the reference frame rotating with the surface at the angular velocity $\bm{\Omega}_o= \Omega_o \bm{e}_z$, the Boussinesq MHD equations read
\begin{align} \label{eq:1}
    \mathbf{\nabla \cdot v} &=0\,,\\ \label{eq:2}
        D_t\mathbf{v} &= -\frac{1}{\rho}{\nabla p'} -2\Omega_o \mathbf{e}_z\times \mathbf{v}+\alpha g T' \mathbf{e}_r 
         + \frac{1}{4 \pi \rho} (\nabla\times \mathbf{B})\times \mathbf{B} + \nu\Delta\mathbf{v}\,, 
    \\ \label{eq:3}
    D_tT' &=\kappa\Delta T'\,,\\ \label{eq:4}
    \partial_t\mathbf{B} &=\nabla\times(\mathbf{u}\times\mathbf{B})+\eta\Delta \mathbf{B}\,,\\ \label{eq:5}
    \nabla\cdot\mathbf{B} &=0\,, 
\end{align}
where $\mathbf{v}$ is the velocity field, $\mathbf{B}$ is the magnetic field, $p'$ is the pressure perturbation, $T'$ is the super-adiabatic temperature, $\rho$ is the uniform density, $g=g_o r/r_o$ is the gravitation field, and $\alpha\equiv \rho^{-1}(\partial_{T}\rho)_{p}$ is the thermal expansion coefficient. $\mathbf{e}_z$ and $\mathbf{e}_r$ are the unit vectors of the axial and the spherical radial directions, respectively. We apply no-slip, electrically insulating, and  fixed temperature boundary conditions on both shells. In the above equations, we assume that the viscosity $\nu$, the thermal diffusivity $\kappa$ and the magnetic diffusivity $\eta$ are constant. Apart from the magnetic diffusivity which relates to the electrical conductivity of electrons, the physical interpretation of the other transport coefficients can lead to different estimates, depending on whether neutrinos are considered or not to be the main source of diffusive processes.

\subsection{Transport coefficients}
In our magnetar formation scenario, the fallback occurs seconds to minutes after the PNS formation. As discussed in \citet[][Sect. 4.3.]{barrere2022}, at this stage of the PNS evolution, the diffusive processes are either microscopic or driven by neutrinos depending on whether the length scale $l$ considered is smaller or larger than the neutrino mean free path \citep[][Eq. 11]{thompson1993}
\begin{equation}
    l_{n}\sim\SI{4e4}{}\left(\frac{\rho}{\SI{e14}{g.cm^{-3}}}\right)^{-1/3}\left(\frac{T}{\SI{5}{MeV}}\right)^{-3}\SI{}{cm}\,.
\end{equation}
If the length scale satisfies $l>l_n$, then the viscosity and the thermal diffusivity are driven by neutrinos. Using the scalings of \citet[][Eq.~10]{guilet2015b} and \citet[][Eq.~7]{thompson1993}, we obtain the following orders of magnitude
\begin{align}
    \nu_{n}&\sim\SI{2e6}{}\left(\frac{\rho}{\SI{4e14}{g.cm^{-3}}}\right)^{-2}\left(\frac{T}{\SI{5}{MeV}}\right)^{2}\SI{}{cm^2.s^{-1}}\,,\\
    \kappa_{n}&\sim\SI{4e10}{}\left(\frac{\rho}{\SI{4e14}{g.cm^{-3}}}\right)^{-2/3}\left(\frac{T}{\SI{5}{MeV}}\right)^{-1}\SI{}{cm^2.s^{-1}}\,.
\end{align}
If the length scale satisfies $l<l_n$, it is relevant to use a microscopic viscosity such as the shear viscosity due to neutron-neutron scattering~\citep[][Eq.~14]{cutler1987}
\begin{equation}
    \nu_{s}\sim0.2\left(\frac{\rho}{\SI{4e14}{g.cm^{-3}}}\right)^{5/4}\left(\frac{T}{\SI{5}{MeV}}\right)^{-2}\SI{}{cm^2.s^{-1}}\,.
\end{equation}

To determine the thermal diffusivity, we use the calculation of \citet[][Eqs. 48--54]{lee1950} for degenerate relativistic electrons

\begin{equation}\label{eq:cond}
    \kappa_s=\frac{\lambda}{\rho c_p}= \frac{8\pi^2cT}{3\rho c_p}\left(\frac{3\pi^2\rho Y_e}{m_p}\right)^{1/3}\left(\frac{k_B}{e}\right)^2\frac{1}{4\pi\alpha\ln \Lambda}\,,
\end{equation}
where $\lambda$ is the thermal conductivity, $c_p$ is the specific heat at constant pressure, $Y_e$ is the electronic fraction, and $c$ is the speed of light in the vacuum. The constants $k_B$, $e$, $m_p$, and $\alpha\equiv e^2/(\hbar c)$ are the Boltzman constant, the electric charge, the proton mass, and the fine structure constant, respectively, and $\ln \Lambda\sim 1$ is the Coulomb logarithm.
By using the formula of the specific heat for noninteracting gas of semidegenerate nucleons of \citet{thompson1993}
\begin{equation}\label{eq:spe_heat}
    \rho c_p=\left(\frac{\pi}{3}\right)^{2/3}\left(\frac{k_B}{\hbar}\right)^2\left[f(Y_e)\right]^{-1}\rho^{1/3}m_p^{2/3}T\,,
\end{equation}
where $f(Y_e)\equiv \left[(1-Y_e)^{1/3}+Y_e^{1/3}\right]^{-1}$, Eq.~\eqref{eq:cond} simplifies in 
\begin{align}
    \kappa_s&=2\pi\frac{c}{m_p\alpha \ln \Lambda}\left(\frac{\hbar}{e}\right)^2\left[f(Y_e)\right] Y_e^{1/3}\\
    &\sim \SI{30}{}\left(\frac{Y_e}{0.2}\right)^{1/3}\SI{}{cm^2.s^{-1}}\,.
\end{align}

Assuming that the Wiedemann-Franz law computed by
\citet[Eq. 15]{kelly1973} for degenerate, relativistic electrons
holds, the thermal diffusity $\kappa_s$ and magnetic diffusivity $\eta$ are related by
\begin{equation}
    \kappa_s=\frac{10\pi^3}{3}\frac{T}{\rho c_p}\left(\frac{k_B}{e}\right)^2\frac{c^2}{4\pi\eta}\,.
\end{equation}
Therefore, the magnetic diffusitivity scales like
\begin{equation}
    \eta=\SI{2e-5}{}\left(\frac{\rho}{\SI{4e14}{g.cm^{-3}}}\right)^{-1/3}\left(\frac{Y_e}{0.2}\right)^{-1/3}\SI{}{cm^2.s^{-1}}\,.
\end{equation}

\subsection{Parameter regime}
The above calculations enable us to estimate the different dimensionless numbers characterising the PNS fluid interior at evolution stage we are interested in. For the thermal and magnetic Prandtl numbers, we have
\begin{equation}
    Pr\equiv \frac{\nu}{\kappa} \sim 
    \left\{
    \begin{aligned}
        \SI{5e-5}{}  &\quad {\rm with} \quad \nu=\nu_n, \kappa=\kappa_n \\
        \SI{7e-3}{} &\quad{\rm with}\quad \nu=\nu_s, \kappa=\kappa_s
    \end{aligned}
    \right.\,,
\end{equation}
and 
\begin{equation}
    Pm\equiv \frac{\nu}{\eta} \sim 
    \left\{
    \begin{aligned}
        \SI{e11}{}  &\quad {\rm with} \quad \nu=\nu_n \\
        \SI{e4}{} &\quad{\rm with}\quad \nu=\nu_s
    \end{aligned}
    \right.\,.
\end{equation}
The Ekman number can be estimated as:
\begin{equation}
    E\equiv \frac{\nu}{r^2\Omega} \sim 
    \left\{
    \begin{aligned}
        \SI{2e-9}{}  &\quad {\rm with} \quad \nu=\nu_n \\
        \SI{2e-16}{} &\quad{\rm with}\quad \nu=\nu_s
    \end{aligned}
    \right.\,,
\end{equation}
with $r=\SI{12}{km}$ and $\Omega=200\pi\,\SI{}{rad.s^{-1}}$.
Finally, the stratification of the PNS interior can be characterised by the Brunt-Väisälä frequency
\begin{equation}\label{eq:brunt-vaisala}
    N\equiv \sqrt{-\frac{g}{\rho}\left(\left.\frac{\partial\rho}{\partial S}\right|_{P,Y_e}\frac{dS}{dr}+\left.\frac{\partial \rho}{\partial Y_e}\right|_{P,S}\frac{dY_e}{dr}\right)}
    \sim\SI{4e3}{s^{-1}}
    \,,
\end{equation}
where $S$ is the entropy. The above order of magnitude for $N$ is based on the 1D core-collapse supernova simulations from~\citet[][Chap.~5]{hudepohl2014}.

In all cases, these parameters are far beyond the reach of any modern supercomputer.
To limit the computational time needed to complete our parameter study, we considered the following values 
$Pr=0.1$, $Pm=1$, $E=\num{e-5}$ and $N/\Omega_o = 0.1$. We leave for future work the study of the dependence on the diffusivity coefficients and strength of stratification.

\subsection{Numerical methods}\label{sec:numerical}

To satisfy the solenoidal conditions \eqref{eq:1} and \eqref{eq:5}, the velocity and magnetic fields are decomposed in poloidal and toroidal components (Mie representation),
\begin{align}
    \rho\mathbf{u}&=\nabla\times(\nabla\times W\mathbf{e}_r)+\nabla\times Z\mathbf{e}_r\,,\\
    \mathbf{B}&=\nabla\times(\nabla\times b\mathbf{e}_r)+\nabla\times a_j\mathbf{e}_r\,,
\end{align}
where W and Z ($b$ and $a_j$) are the poloidal and toroidal potentials for the velocity (magnetic) field. The whole system of equations is then solved in spherical coordinates by expanding the scalar potentials in Chebyshev polynomials in the radial direction, and in spherical harmonic functions in the angular directions. We refer the reader to the MagIC online documentation\footnote{\url{https://magic-sph.github.io}} for an exhaustive presentation of the numerical techniques \citep[see also][]{wicht2002,gastine2012,schaeffer2013}.

\subsection{Output parameters}

We first characterize our models by computing the time average of the kinetic and magnetic energy densities (after filtering out any initial transient). The latter is expressed in terms of the Elsasser number $\Lambda\equiv B_{\rm rms}^2/(4\pi\rho\eta\Omega_o)$ and used to compute different rms estimates of the magnetic field. In addition to the total field, we distinguish the poloidal and toroidal fields based on the Mie representation (Sect.~\ref{sec:numerical}), while the dipole field refers to the $l=1$ poloidal component.

\section{Supplemental outputs}

\subsection{Force balance}\label{sec:force}
\begin{figure*}
    \centering
    \includegraphics[width=0.95\textwidth]{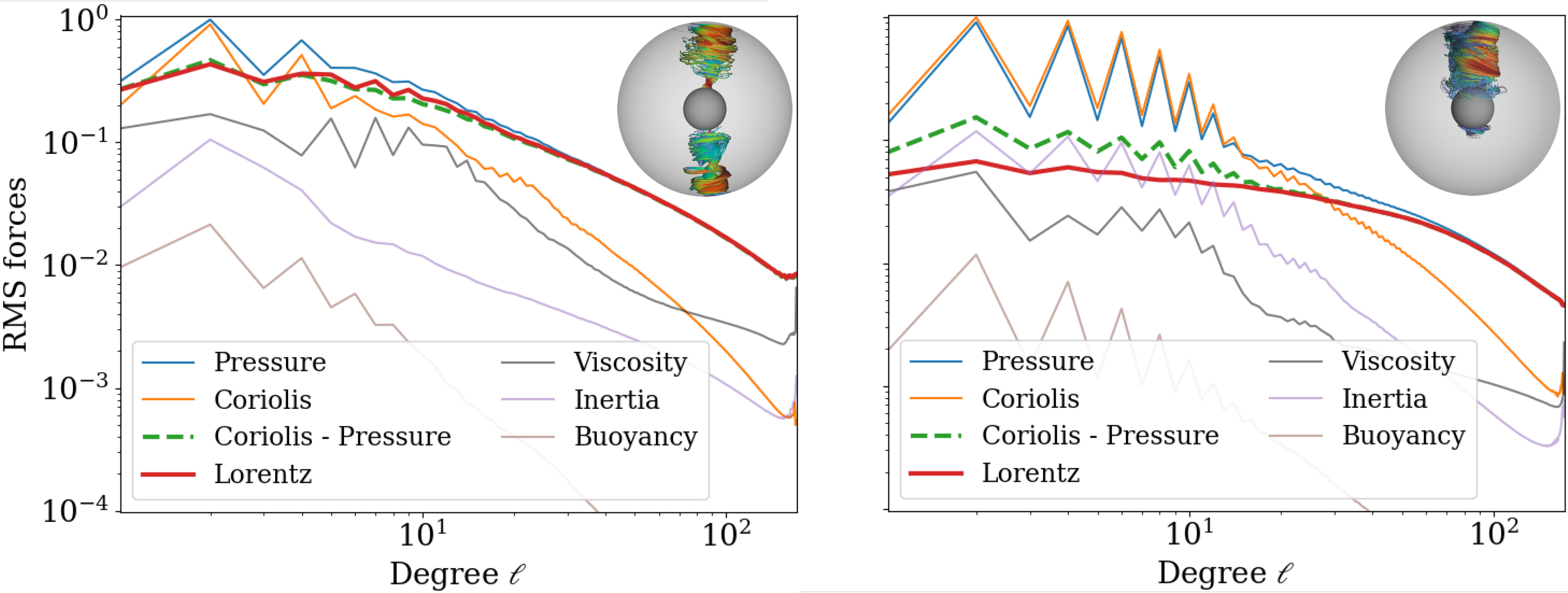}
    \caption{Time-averaged rms force spectra for the dipolar (left) and hemispherical (right) dynamos at $Ro=0.75$. The rms forces are averaged over the whole computational domain without excluding boundary layers.}
    \label{fig:force_balance}
\end{figure*}

Fig.~\ref{fig:force_balance} shows a spectrum of the rms forces in the saturated state of the two Tayler-Spruit dynamo branches, following the formalism of \citet{aubert2017,schwaiger2019}. The dipolar dynamo saturates due to a balance between the Lorentz force (red line) and the ageostrophic Coriolis force (dashed green line) at all scales (spectrum on the left in Fig.~\ref{fig:force_balance}). This confirms the magnetostrophic balance we deduced from Fig.~2 in the Letter. For the hemispherical dynamo, the same balance is found at small scales ($\ell\gtrsim20$), but at large scales the inertial force is strong enough to be in balance with the ageostrophic Coriolis force and the Lorentz force.

\subsection{Time-averaged spectra}\label{sec:spec}
\begin{figure*}
    \centering
    \includegraphics[width=0.45\textwidth]{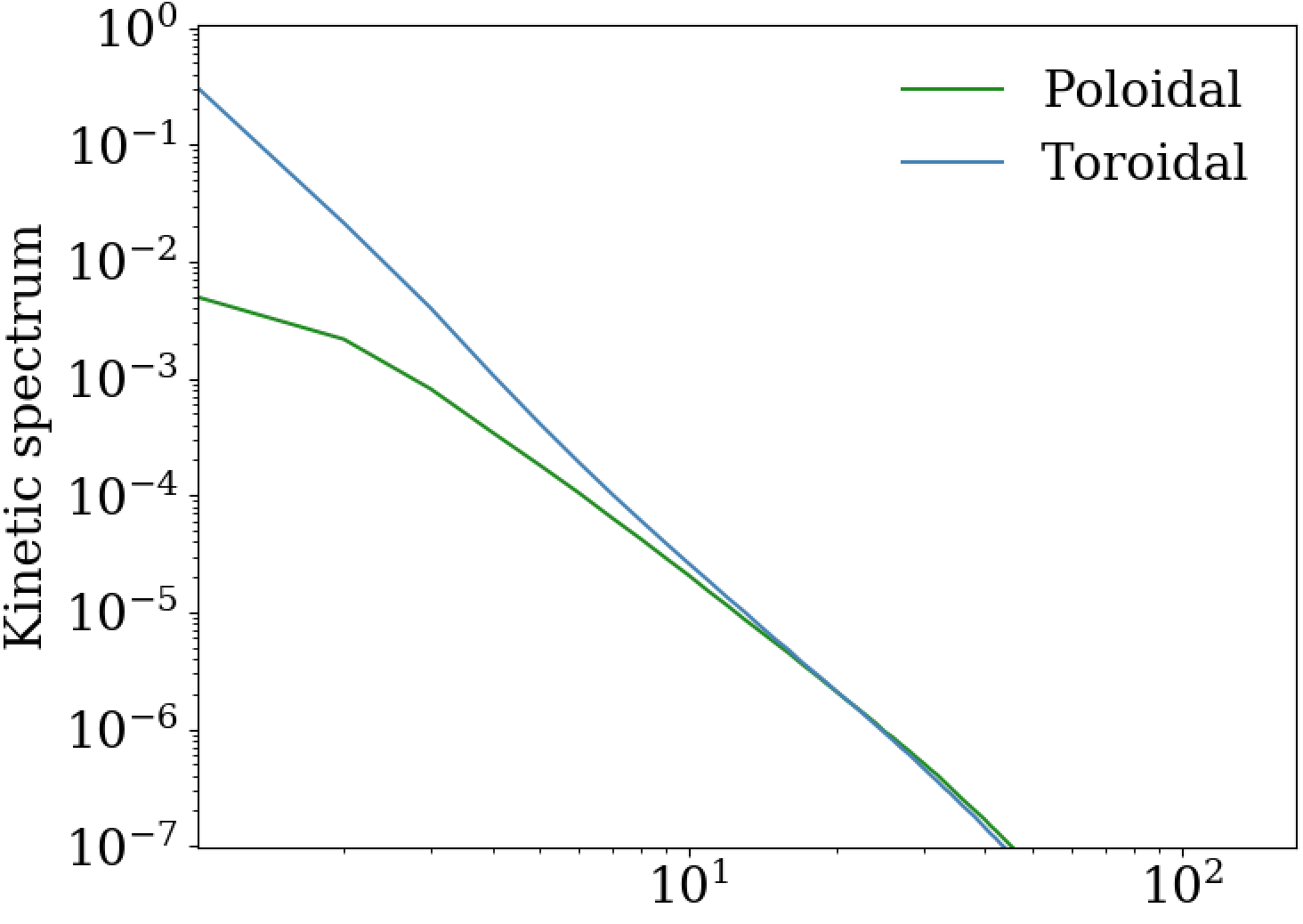}
    \includegraphics[width=0.44\textwidth]{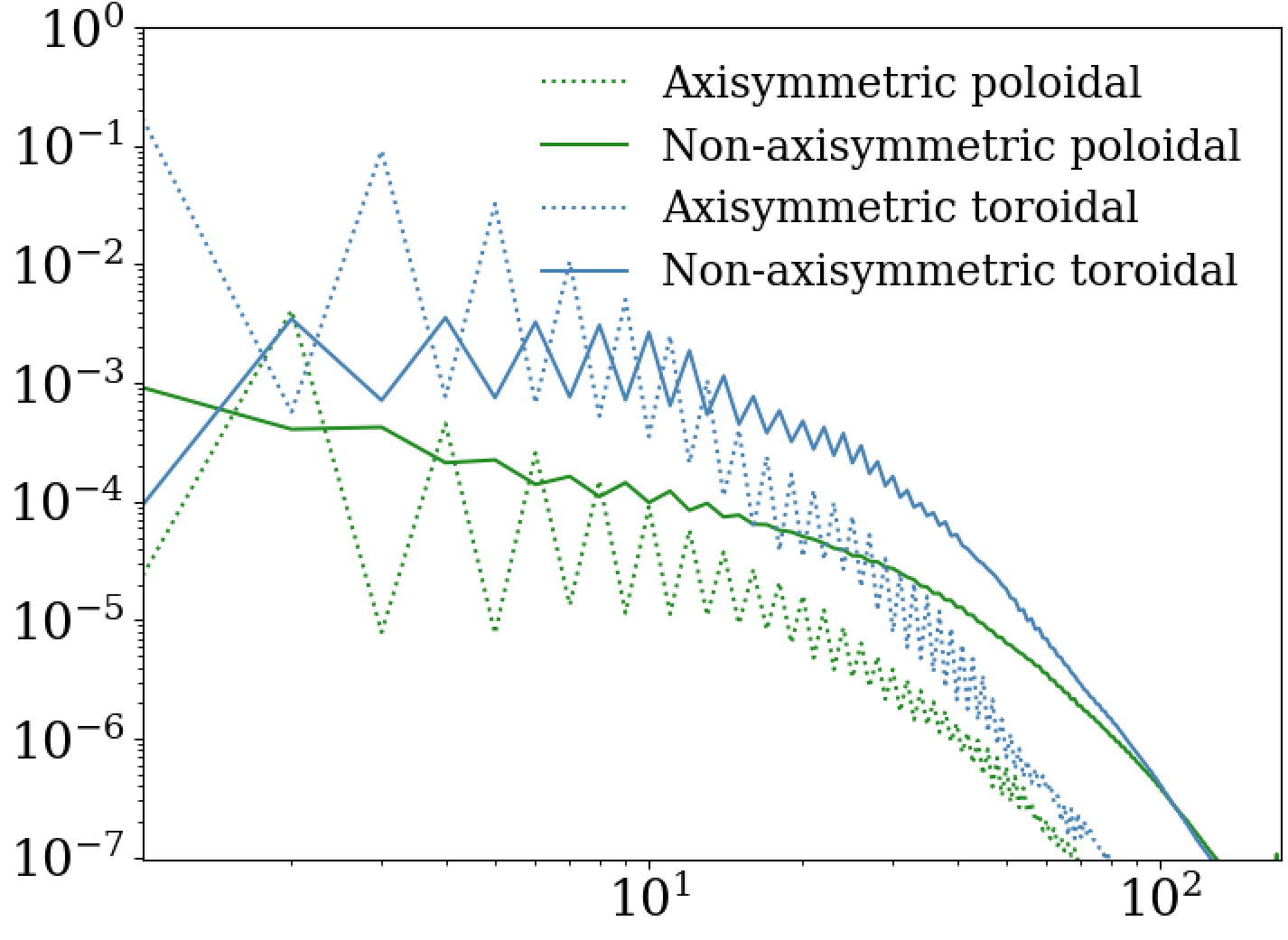}
    \includegraphics[width=0.45\textwidth]{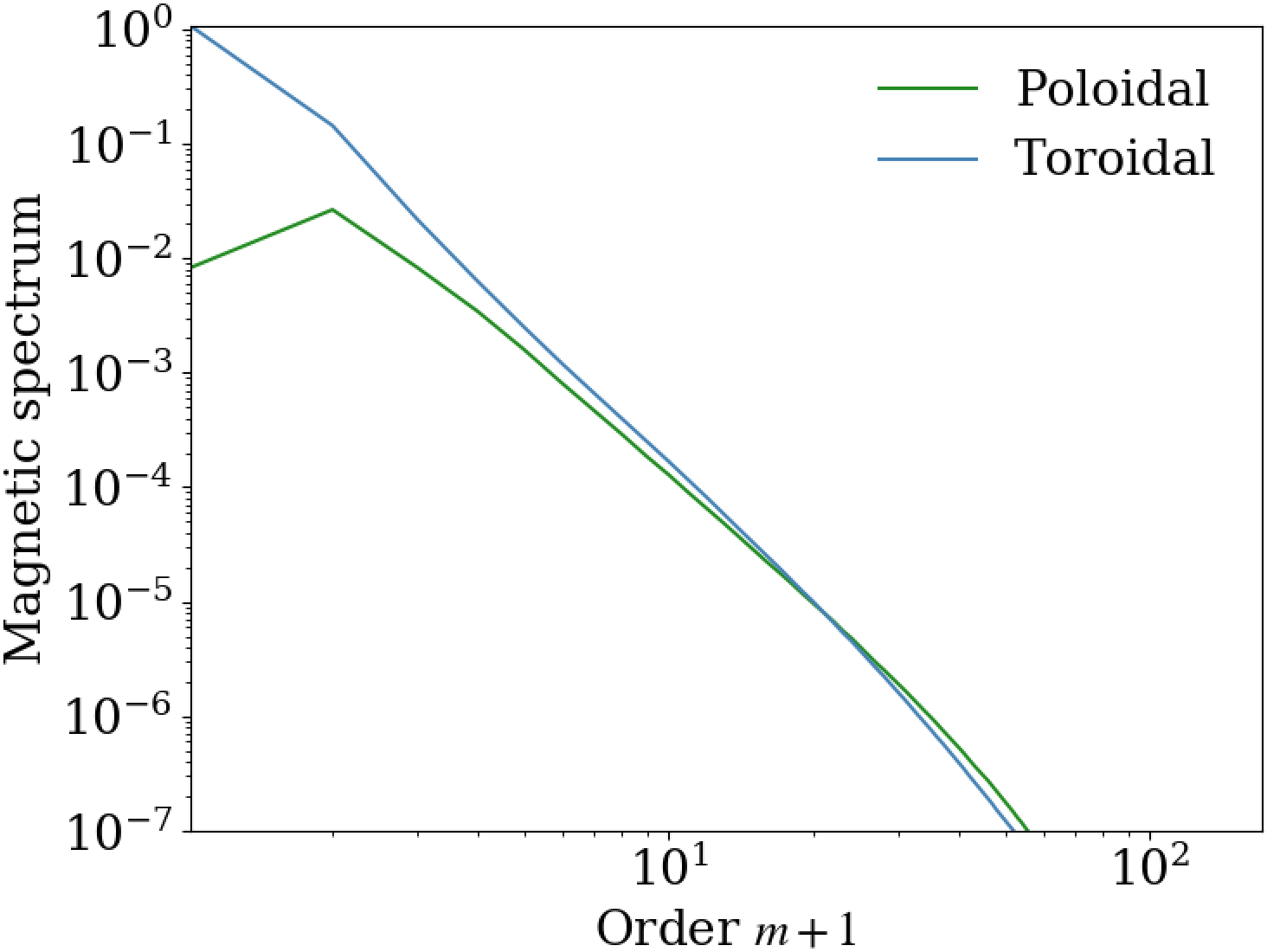}
    \includegraphics[width=0.44\textwidth]{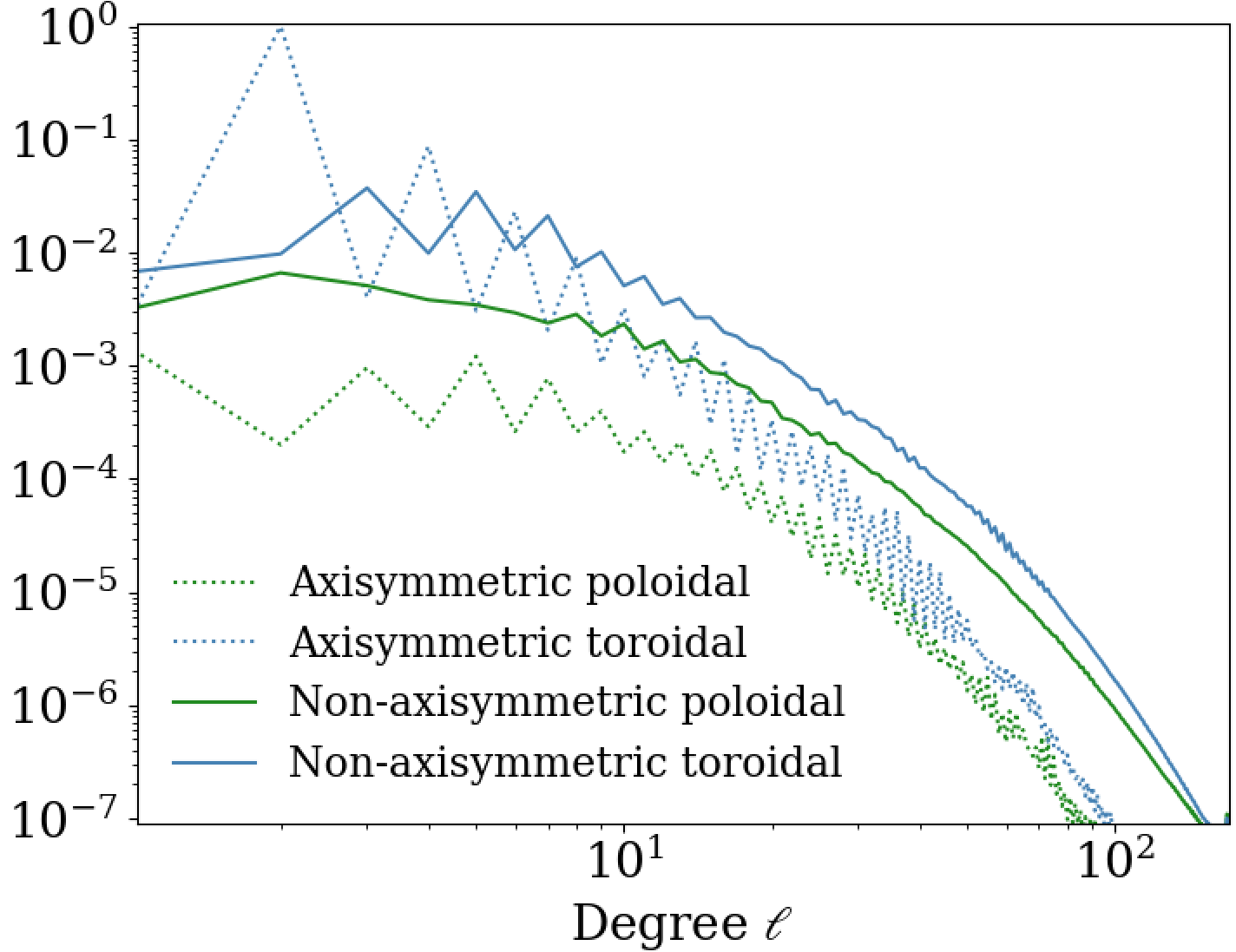}
    \caption{Time averaged kinetic (top) and magnetic (bottom) energy density spectra of the dipolar Tayler-Spruit dynamo at $Ro=0.75$. 
    }
    \label{fig:spectra}
\end{figure*}

A wide range of modes $\ell$ are present in the typical spectra of a dipolar Tayler-Spruit dynamo of Fig.~\ref{fig:spectra}. The magnetic spectrum shows the presence of a significant large-scale axisymmetric poloidal field. The even (odd) degrees~$\ell$ dominate in the poloidal (toroidal) axisymmetric magnetic spectra, which confirms the dipolar equatorial symmetry at large scales. 

\begin{figure*}
    \centering
    \includegraphics[width=0.95\textwidth]{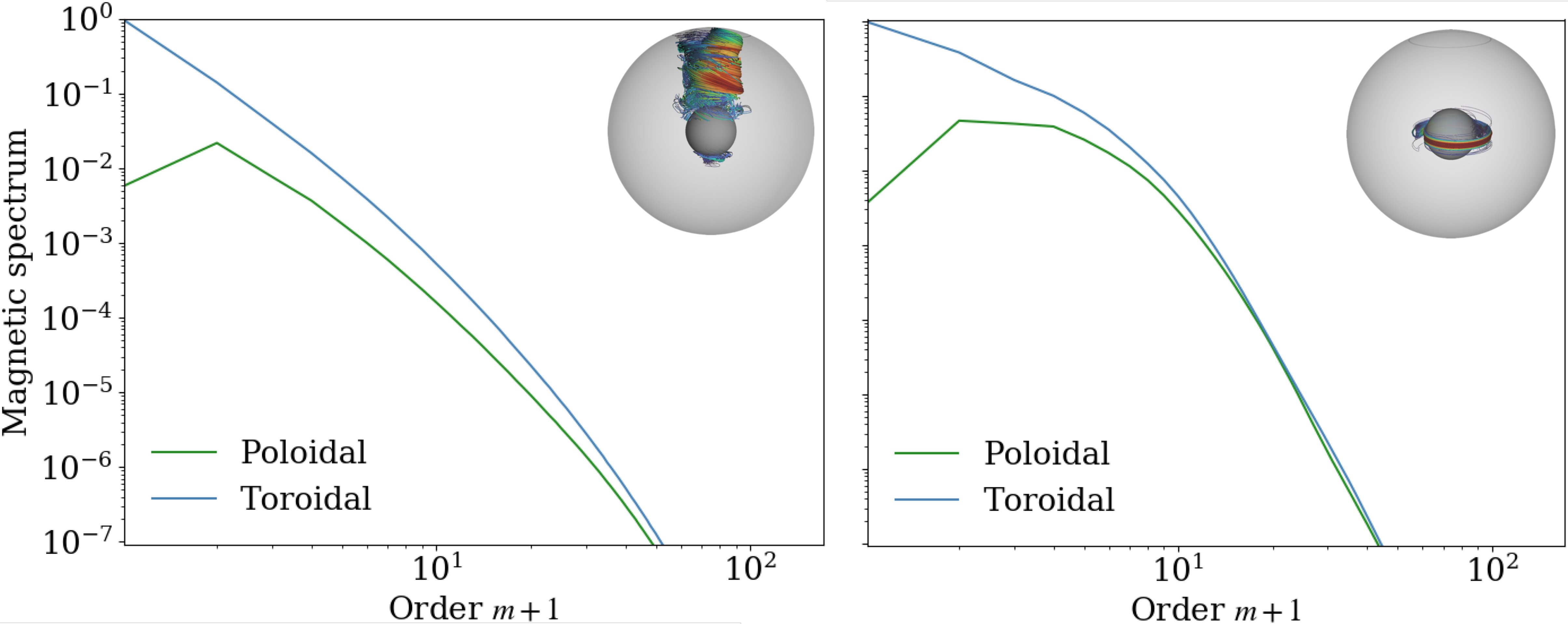}
    \caption{Time averaged $m$-spectra of the magnetic energy density for the hemispherical Tayler-Spruit dynamo at $Ro=0.85$ (left) and the kinematic dynamo at $Ro=0.75$ (right).}
    \label{fig:specm}
\end{figure*}

The non-axisymmetric modes triggered differ depending on whether the dynamo is Tayler-instability driven, as seen in Figures~\ref{fig:spectra} and \ref{fig:specm}. As expected for the Tayler-Spruit dynamo, the dominating mode is the axisymmetric toroidal magnetic field, but we also observe a dominant $m=1$ mode in the poloidal magnetic energy, which is a signature of the Tayler instability. By contrast, we see that a wider range of orders $m\in[1,5]$ are present in the poloidal magnetic energy of the kinematic dynamo.

\subsection{Geometrical criterion for Tayler instability}\label{sec:Tayler}

\begin{figure*}
    \centering
    \includegraphics[width=0.75\textwidth]{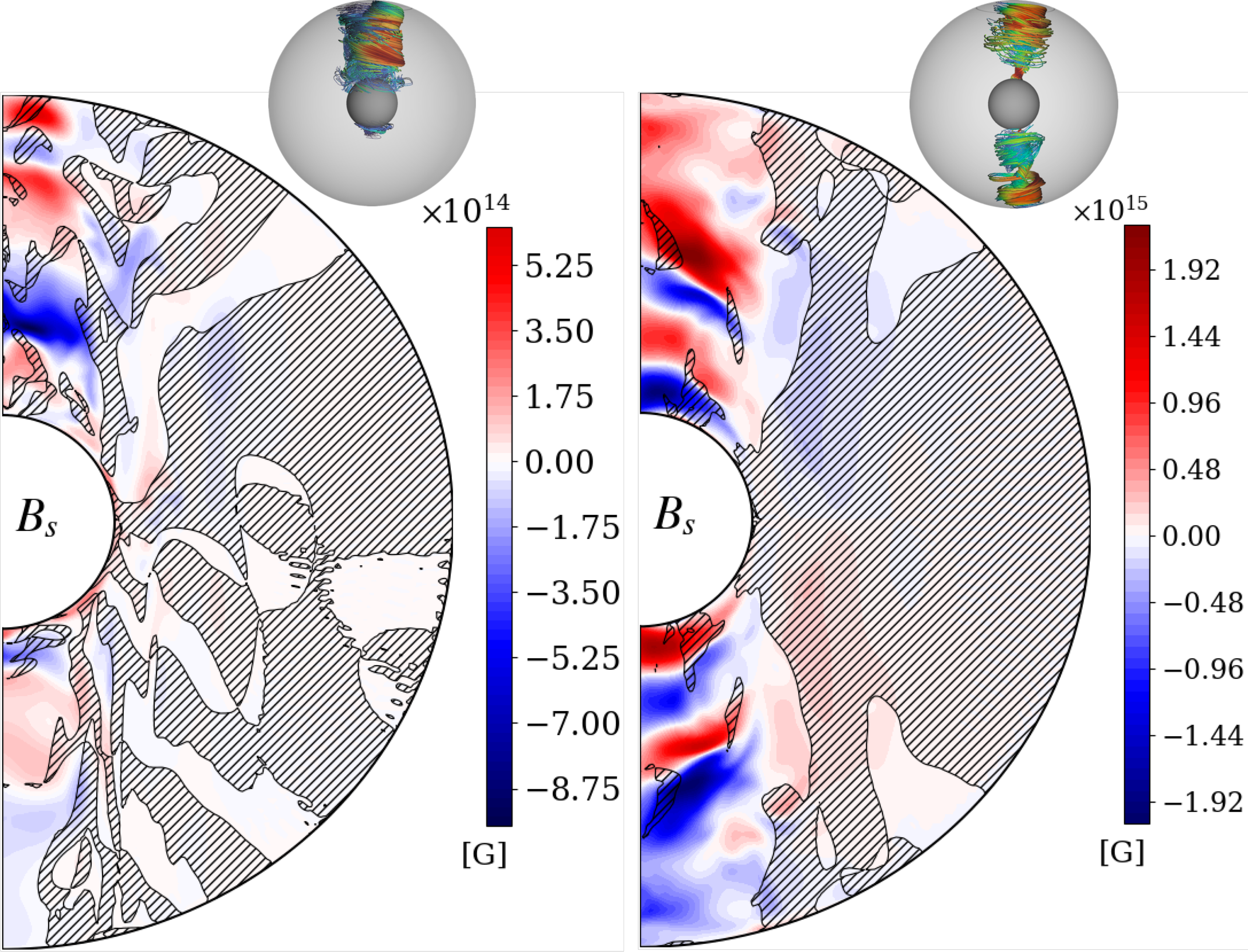}
    \caption{Snapshots of the meridional slices of $B_s$ in the runs Ro0.75w and Ro0.75s. The hatched areas correspond to the regions where the fluid is stable to $m=1$ perturbations ($c_{m=1}>0$).}
    \label{fig:goossens}
\end{figure*}

\citet{tayler1973} showed that any axisymmetric toroidal field is unstable to adiabatic perturbations in ideal MHD (i.e. $\nu=\eta=0$) in a non-rotating stratified  fluid and worked out the necessary and sufficient conditions for magnetic instability (now referred to as \emph{Tayler instability}). 
In spherical coordinates, they read \citep{goossens1980b}
\begin{equation}
    c_{m=0} \equiv\frac{B_{\phi}^2}{2\pi r^2 \sin^2\theta}\left(\cos^2\theta-\sin\theta\cos\theta\partial_{\theta}\log B_{\phi}\right)< 0 \,,
\end{equation}
for axisymmetric perturbations,  and
\begin{equation}
    c_{m\neq 0} \equiv\frac{B_{\phi}^2}{4\pi r^2 \sin^2\theta}\left(m^2-2\cos^2\theta-2\sin\theta\cos\theta\partial_{\theta}\log B_{\phi}\right)< 0 \,,
\end{equation}
for non-axisymmetric perturbations.
Since the $m=1$ mode is the most unstable (which is consistent with our simulations), we display in Fig.~\ref{fig:goossens} the sign of the $c_{m=1}$ coefficient superimposed with the magnetic field for a dipolar and an hemispherical dynamo.  We see that in both cases the perturbed magnetic mainly develops inside the tangent cylinder, which globally matches with the areas that are expected to be unstable to the Tayler instability ($c_{m=1}<0$). This is therefore an additional indication of the presence of the Tayler instability in our simulations.

\subsection{Measure of the shear rate} \label{sec:shear}
\begin{figure*}
    \centering
    \includegraphics[width=0.7\textwidth]{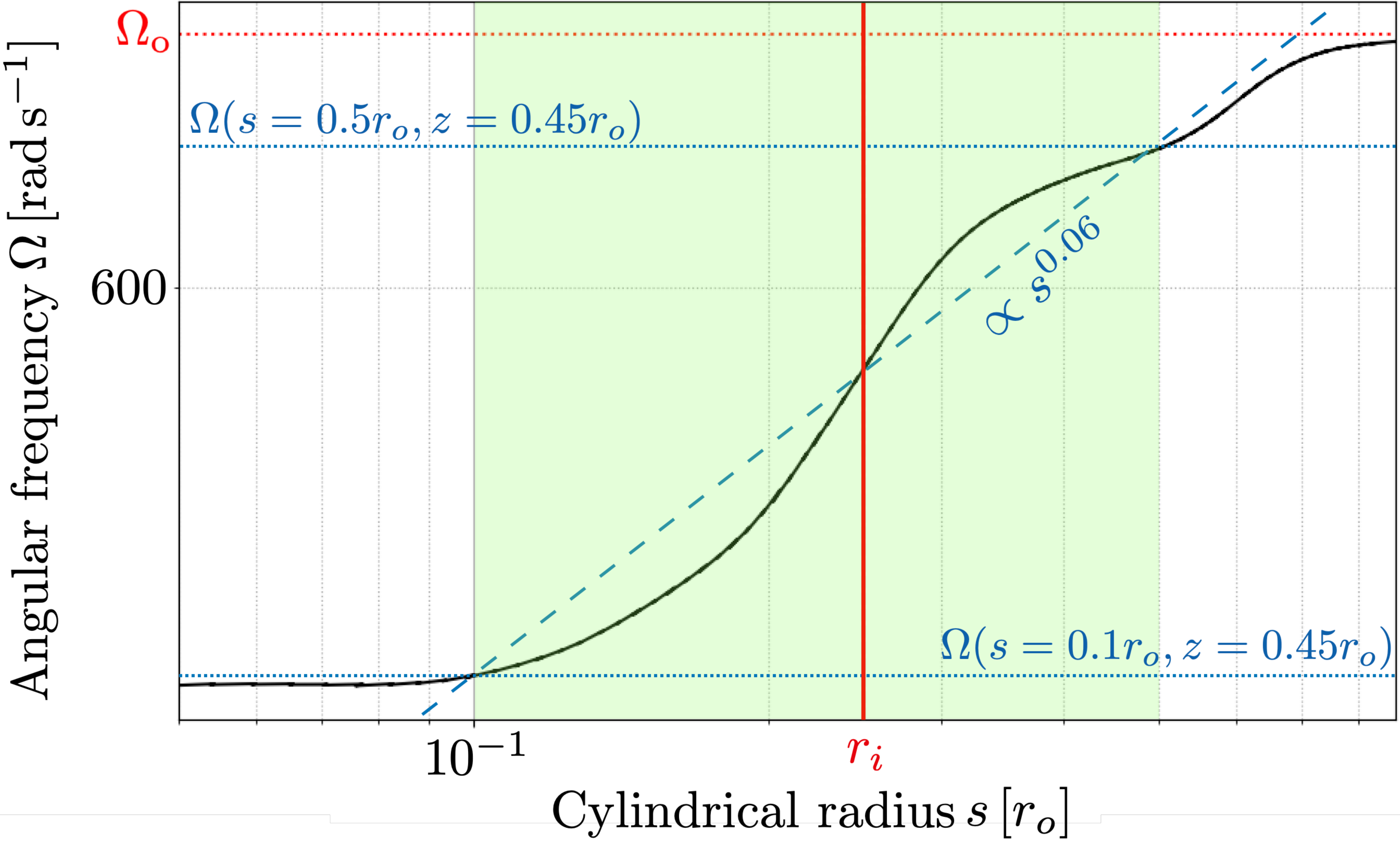}
    \caption{Rotation profile $\Omega(s)$ at $z=0.45r_o$ in the simulation {Ro0.75s}. The green region $s\in[0.1,0.5]\,r_o$ is the zone where we measure the effective shear rate~$q$ (slope of the blue dashed line). In this example, $q\sim 0.06$. The vertical red line indicates the position of the tangent cylinder.}
    \label{fig:cyl_prof}
\end{figure*}

The differential rotation is characterized by a dimensionless shear rate $q=r\partial_r {\rm ln}{\Omega}$. We define an effective shear rate based on the time average of the rotation profile in the saturated state. Since it is approximately cylindrical (see Fig.~3), we measure $\Omega$ locally at a given height $z=0.45r_o$ and consider its variation as a function of the cylindrical radius $s$ (see Fig.~\ref{fig:cyl_prof}). This allows us to avoid the Ekman layers, which form around the inner shell. In most of the simulations, the shear is found in a broad region centred on the tangent cylinder, especially in $s\in[0.1,0.5]\,r_o$. We therefore measure the average slope of the profile in this interval (see Fig.~\ref{fig:cyl_prof}):

\begin{equation}
    q\equiv\frac{\log\Omega(s=0.5r_o)-\log\Omega(s=0.1r_o)}{\log0.5r_o-\log0.1r_o}
\end{equation}

\newpage
\section{List of models}\label{sec:simu}

Tables~\ref{tab:sim_dynamo} and~\ref{tab:sim_trans} summarize the key parameters of the simulations carried out in this study.

\onecolumn{
\begin{landscape}
\begin{table*}
\caption{Overview of the stable (or failed) dynamos solutions. 
The simulations are named using the value of the Rossby number followed by a letter which characterises the initial condition: `s' for a strong dipole-like magnetic field (either analytic or from a saturated state of the dipolar branch), `as' for a strong quadrupole-like field (analytic or from a saturated state of the quadrupolar branch), `h' for the saturated state of the hemispherical branch, `k' for a saturated state of the kinematic dynamo, and `w' for a very weak seed field. The second column indicates the input Rossby number. The third column refers to the initial Elsasser number. The last column specifies the type of the dynamo solution: dipolar (DIP), quadrupolar (QUAD), hemispherical (HEM), oscillating (OSC), kinematic (WEAK), non dynamo (NONE). All the simulations have the same aspect ratio $\chi=0.25$, Ekman number $E=10^{-5}$, stratification $N/\Omega_o=0.1$, thermal and magnetic Prandtl numbers $Pr=0.1$ and $Pm=1$, and the same resolution $(n_r,n_{\theta},n_{\phi})=(256,256,512)$.}
\begin{tabular}{lccccccccccccr}
\hline
 Name & ${Ro}$ & ${\Lambda_i}$ & ${q}$ & $\Lambda$ & $E_b/E_k$ & $B_{\mathrm{tor}}^{m=0}$ & $B_{\mathrm{pol}}^{m=0}$ & $B_{\mathrm{dip}}$ & $B_sB_{\phi}/4\pi$ & $B_rB_{\phi}/4\pi$ & $B_s^{m=0}B_{\phi}^{m=0}/4\pi$ & $B_r^{m=0}B_{\phi}^{m=0}/4\pi$ & Dynamo \\
   &   &   &   &   &   & $[10^{14}\,{\rm G}]$ & $[10^{14}\,{\rm G}]$ & $[10^{14}\,{\rm G}]$ & $[10^{27}\,{\rm G}^2]$ & $[10^{27}\,{\rm G}^2]$ & $[10^{27}\,{\rm G}^2]$ & $[10^{27}\,{\rm G}^2]$ &   \\
\hline
$ \text{Ro1.2s} $ & $1.2$ & $ \text{$\Lambda(\mathrm{Ro1s})$} $ & $0.13$ & $70.11$ & $1.73$ & $9.16$ & $1.04$ & $0.70$ & $41.97$ & $45.70$ & $8.31$ & $25.11$ & $ \text{DIP} $ \\
$ \text{Ro1s} $ & $1$ & $ \text{$\Lambda(\mathrm{Ro0.75s})$} $ & $0.069$ & $87.83$ & $4.74$ & $10.46$ & $1.17$ & $0.77$ & $49.37$ & $53.33$ & $10.80$ & $30.60$ & $ \text{DIP} $ \\
$ \text{Ro0.75s} $ & $0.75$ & $ \text{$10$} $ & $0.06$ & $51.85$ & $3.93$ & $8.08$ & $0.68$ & $0.48$ & $22.57$ & $21.10$ & $4.43$ & $10.24$ & $ \text{DIP} $ \\
$ \text{Ro0.5s} $ & $0.5$ & $ \text{$\Lambda(\mathrm{Ro0.75s})$} $ & $0.032$ & $33.06$ & $6.16$ & $6.65$ & $0.46$ & $0.33$ & $11.31$ & $9.84$ & $2.20$ & $4.02$ & $ \text{DIP} $ \\
$ \text{Ro0.35s} $ & $0.35$ & $ \text{$\Lambda(\mathrm{Ro0.25s})$} $ & $0.02$ & $20.87$ & $8.36$ & $5.34$ & $0.39$ & $0.25$ & $7.01$ & $6.39$ & $1.53$ & $2.80$ & $ \text{DIP} $ \\
$ \text{Ro0.2s} $ & $0.2$ & $ \text{$\Lambda(\mathrm{Ro0.25s})$} $ & $0.0089$ & $10.16$ & $11.67$ & $3.70$ & $0.23$ & $0.15$ & $3.22$ & $2.91$ & $0.56$ & $1.27$ & $ \text{DIP} $ \\
$ \text{Ro0.17s} $ & $0.17$ & $ \text{$\Lambda(\mathrm{Ro0.25s})$} $ & -- & -- & -- & -- & -- & -- & -- & -- & -- & -- & $ \text{NONE} $ \\
$ \text{Ro0.125s} $ & $0.125$ & $ \text{$\Lambda(\mathrm{Ro0.25s})$} $ & -- & -- & -- & -- & -- & -- & -- & -- & -- & -- & $ \text{NONE} $ \\
\hline
$ \text{Ro1as} $ & $1$ & $ \text{$\Lambda(\mathrm{Ro0.75as})$} $ & $0.45$ & $13.75$ & $0.19$ & $3.83$ & $0.33$ & $0.32$ & $9.28$ & $7.64$ & $1.68$ & $2.66$ & $ \text{HEM} $ \\
$ \text{Ro0.75as} $ & $0.75$ & $ \text{$10$} $ & $0.064$ & $58.22$ & $4.46$ & $8.71$ & $0.67$ & $0.58$ & $21.77$ & $20.47$ & $4.18$ & $10.01$ & $ \text{DIP} $ \\
$ \text{Ro0.65as} $ & $0.65$ & $ \text{$\Lambda(\mathrm{Ro0.75as})$} $ & $0.27$ & $5.15$ & $0.13$ & $2.34$ & $0.19$ & $0.19$ & $3.94$ & $3.13$ & $0.62$ & $0.93$ & $ \text{HEM} $ \\
$ \text{Ro0.5as} $ & $0.5$ & $ \text{$\Lambda(\mathrm{Ro0.75as})$} $ & $0.2$ & $2.32$ & $0.09$ & $1.50$ & $0.11$ & $0.17$ & $2.49$ & $1.73$ & $0.31$ & $0.33$ & $ \text{HEM} $ \\
$ \text{Ro0.4as} $ & $0.4$ & $ \text{$\Lambda(\mathrm{Ro0.75as})$} $ & $0.16$ & $0.28$ & $0.01$ & $0.53$ & $0.02$ & $0.04$ & $0.35$ & $0.27$ & $0.02$ & $0.04$ & $ \text{HEM} $ \\
\hline
$ \text{Ro1.2h} $ & $1.2$ & $ \text{$\Lambda(\mathrm{Ro1k})$} $ & $0.51$ & $41.35$ & $0.45$ & $7.28$ & $0.56$ & $0.43$ & $18.27$ & $16.40$ & $3.98$ & $6.90$ & $ \text{OSC} $ \\
$ \text{Ro0.42h} $ & $0.42$ & $ \text{$\Lambda(\mathrm{Ro0.5k})$} $ & $0.17$ & $0.49$ & $0.02$ & $0.63$ & $0.03$ & $0.09$ & $0.89$ & $0.70$ & $0.06$ & $0.07$ & $ \text{HEM} $ \\
\hline
$ \text{Ro1k} $ & $1$ & $ \text{$\Lambda(\mathrm{Ro0.75w})$} $ & $0.43$ & $16.68$ & $0.24$ & $4.37$ & $0.34$ & -- & $8.71$ & $7.30$ & $1.45$ & $2.62$ & $ \text{OSC} $ \\
$ \text{Ro0.5k} $ & $0.5$ & $ \text{$\Lambda(\mathrm{Ro0.75w})$} $ & $0.2$ & $1.46$ & $0.06$ & $1.06$ & $0.08$ & $0.14$ & $1.96$ & $1.35$ & $0.19$ & $0.22$ & $ \text{HEM} $ \\
\hline
$ \text{Ro0.85w} $ & $0.85$ & $ \text{$10^{-4}$} $ & $0.33$ & $14.32$ & $0.26$ & $4.07$ & $0.31$ & $0.36$ & $7.58$ & $6.56$ & $1.41$ & $2.57$ & $ \text{OSC} $ \\
$ \text{Ro0.75w} $ & $0.75$ & $ \text{$10^{-4}$} $ & $0.35$ & $5.12$ & $0.11$ & $2.13$ & $0.20$ & $0.24$ & $4.59$ & $3.50$ & $0.65$ & $1.00$ & $ \text{HEM} $ \\
$ \text{Ro0.75w2} $ & $0.75$ & $ \text{$10^{-4}$} $ & $0.34$ & $0.37$ & $0.01$ & $0.52$ & $0.03$ & $0.08$ & $1.44$ & $1.26$ & $0.06$ & $0.06$ & $ \text{KIN} $ \\
$ \text{Ro0.65w} $ & $0.65$ & $ \text{$10^{-4}$} $ & $0.27$ & $4.87$ & $0.13$ & $2.16$ & $0.18$ & -- & $3.93$ & $3.11$ & $0.60$ & $0.87$ & $ \text{HEM} $ \\
$ \text{Ro0.63w} $ & $0.63$ & $ \text{$10^{-4}$} $ & -- & -- & -- & -- & -- & -- & -- & -- & -- & -- & $ \text{NONE} $ \\
$ \text{Ro0.6w} $ & $0.6$ & $ \text{$10^{-4}$} $ & -- & -- & -- & -- & -- & -- & -- & -- & -- & -- & $ \text{NONE} $ \\
$ \text{Ro0.55w} $ & $0.55$ & $ \text{$10^{-4}$} $ & -- & -- & -- & -- & -- & -- & -- & -- & -- & -- & $ \text{NONE} $ \\
$ \text{Ro0.35w} $ & $0.35$ & $ \text{$10^{-4}$} $ & -- & -- & -- & -- & -- & -- & -- & -- & -- & -- & $ \text{NONE} $ \\
\hline
\end{tabular}
\label{tab:sim_dynamo}
\end{table*}
\begin{table*}
\caption{Same as Table~\ref{tab:sim_dynamo} but for the metastable solutions.}
\begin{tabular}{lccccccccccccr}
\hline
 Name & ${Ro}$ & ${\Lambda_i}$ & ${q}$ & $\Lambda$ & $E_b/E_k$ & $B_{\mathrm{tor}}^{m=0}$ & $B_{\mathrm{pol}}^{m=0}$ & $B_{\mathrm{dip}}$ & $B_sB_{\phi}/4\pi$ & $B_rB_{\phi}/4\pi$ & $B_s^{m=0}B_{\phi}^{m=0}/4\pi$ & $B_r^{m=0}B_{\phi}^{m=0}/4\pi$ & Dynamo \\
   &   &   &   &   &   & $[10^{14}\,{\rm G}]$ & $[10^{14}\,{\rm G}]$ & $[10^{14}\,{\rm G}]$ & $[10^{27}\,{\rm G}^2]$ & $[10^{27}\,{\rm G}^2]$ & $[10^{27}\,{\rm G}^2]$ & $[10^{27}\,{\rm G}^2]$ &   \\
\hline
$ \text{Ro1as} $ & $1$ & $ \text{$\Lambda(\mathrm{Ro0.75as})$} $ & $0.45$ & $70.52$ & $1.18$ & $9.71$ & $0.68$ & $0.47$ & -- & -- & -- & -- & $ \text{HEM} $ \\
$ \text{Ro0.75as} $ & $0.75$ & $ \text{$10$} $ & $0.064$ & $50.94$ & $1.77$ & $8.13$ & $0.59$ & $0.51$ & -- & -- & -- & -- & $ \text{DIP} $ \\
$ \text{Ro0.75w} $ & $0.75$ & $ \text{$10^{-4}$} $ & $0.35$ & $0.32$ & $0.01$ & $0.50$ & $0.03$ & $0.08$ & -- & -- & -- & -- & $ \text{HEM} $ \\
$ \text{Ro0.65w} $ & $0.65$ & $ \text{$10^{-4}$} $ & $0.27$ & $0.36$ & $0.01$ & $0.44$ & $0.03$ & -- & -- & -- & -- & -- & $ \text{HEM} $ \\
$ \text{Ro0.25s} $ & $0.25$ & $ \text{$\Lambda(\mathrm{Ro0.5s})$} $ & $0.012$ & $13.60$ & $10.90$ & $4.28$ & $0.26$ & $0.17$ & $4.15$ & $3.59$ & $0.70$ & $1.36$ & $ \text{DIP} $ \\
\hline
\end{tabular}
\label{tab:sim_trans}
\end{table*}
\end{landscape}
}

\bibliographystyle{mnras}
\bibliography{biblio}
\bsp	
\label{lastpage}